\documentclass[manuscript]{emulateapj}

\usepackage{graphics}
\usepackage{amssymb}
\usepackage{color}
\usepackage{enumerate}

\begin{document}

\title{The long lives of giant clumps and the birth of outflows in gas-rich galaxies at high redshift}

\author{Fr\'ed\'eric Bournaud\altaffilmark{1}, 
Valentin Perret\altaffilmark{2}, 
Florent Renaud\altaffilmark{1}, 
Avishai Dekel\altaffilmark{3}, 
Bruce G. Elmegreen\altaffilmark{4}, 
Debra M. Elmegreen\altaffilmark{5}, 
Romain Teyssier\altaffilmark{6}, 
Philippe Amram\altaffilmark{2}, 
Emanuele Daddi\altaffilmark{1}, 
Pierre-Alain Duc\altaffilmark{1}, 
David Elbaz\altaffilmark{1}, 
Benoit Epinat\altaffilmark{2}, 
Jared M. Gabor\altaffilmark{1}, 
St\'ephanie Juneau\altaffilmark{1}, 
Katarina Kraljic\altaffilmark{1},  Emeric Le~Floch'\altaffilmark{1}}

\altaffiltext{1}{CEA, IRFU/SAp, 91191 Gif-Sur-Yvette, France}
\altaffiltext{2}{Aix Marseille Universit\'e, CNRS, LAM (Laboratoire d'Astrophysique de Marseille), 13388, Marseille, France}
\altaffiltext{3}{Center for Astrophysics and Planetary Science, Racah Institute of Physics, The Hebrew University, Jerusalem 91904, Israel}
\altaffiltext{4}{IBM Research Division, T.J. Watson Research Center, Yorktown Heights, NY 10598, USA}
\altaffiltext{5}{Department of Physics and Astronomy, Vassar College, Poughkeepsie, NY 12604, USA}
\altaffiltext{6}{Institute for Theoretical Physics, University of Zurich, CH-8057 Zurich, Switzerland \bigskip}

\begin{abstract}
Star-forming disk galaxies at high redshift are often subject to violent disk instability, characterized by giant clumps whose fate is yet to be understood. The main question is whether the clumps disrupt within their dynamical timescale ($\leq$\,50\,Myr), like the molecular clouds in today's galaxies, or whether they survive stellar feedback for more than a disk orbital time ($\approx$\,300\,Myr) in which case they can migrate inward and help building the central bulge. We present 3.5-7\,pc resolution AMR simulations of high-redshift disks including photo-ionization, radiation pressure, and supernovae feedback. Our modeling of radiation pressure determines the mass loading and initial velocity of winds from basic physical principles. We find that the giant clumps produce steady outflow rates comparable to and sometimes somewhat larger than their star formation rate, with velocities largely sufficient to escape galaxy. The clumps also lose mass, especially old stars, by tidal stripping, and the stellar populations contained in the clumps hence remain relatively young ($\leq$\,200\,Myr), as observed. The clumps survive gaseous outflows and stellar loss, because they are wandering in gas-rich turbulent disks from which they can re-accrete gas at high rates compensating for outflows and tidal stripping, overall keeping realistic and self-regulated gaseous and stellar masses. Our simulations produce gaseous outflows with velocities, densities and mass loading consistent with observations, and at the same time suggest that the giant clumps survive for hundreds of Myr and complete their migration to the center of high-redshift galaxies, without rapid dispersion and reformation of clumps. These long-lived clumps can be involved in inside-out evolution and thickening of the disk, spheroid growth and fueling of the central black hole.
\end{abstract}

%%%%%%%%%%%%%%%%%%%%%%%%%%%%%%%%%%%%%%%%%%%%%%%%%%

\section{Introduction}

Massive high-redshift galaxies often have irregular morphologies dominated by giant clumps in optical imaging surveys \citep{Cowie96,E07}, ionized gas observations \citep{genzel08}, and resolved molecular gas maps \citep{tacconi13}. Typically, a clumpy star-forming galaxy of stellar mass of a few $10^{10-11}$\,M$_\odot$ at redshift $z\,\approx\,1-3$ contains a handful of kpc-sized clumps containing each a few $10^{8-9}$\,M$_{\odot}$ of gas and stars \citep{EE05, genzel11}. About half of the star formation occurs in the giant clumps, the other half being in an apparently diffuse disk probably consisting of smaller, unresolved gas clouds \citep{E09}. At last half of star-forming galaxies at $z\,>\,1$ in the CANDELS near-infrared imaging survey are clumpy (Mozena et al. in preparation, Guo et al. in preparation).

From the following set of evidence, it is now widely accepted that the majority of the clumps form in-situ by gravitational instability with high characteristic Jeans mass and length (around $10^{9}$\,M$_{\odot}$ and 1\,kpc, respectively). The host galaxies apparent shapes range from round to linear, with the respective fractions expected for face-on to edge-on disks \citep[with $\approx$\,1\,kpc scaleheights,][]{E04} and disk-like kinematics with high turbulent velocity dispersions in both $H\alpha$ and CO \citep[][with perhaps some redshift evolution, Puech (2010)]{genzel06,shapiro08, bournaud08, contini12, daddi10co,epinat12, tacconi13}. The high fractions of dense molecular gas (about half of the baryonic mass, Daddi et al. 2008, 2010b, Tacconi et al. 2010) naturally explain the ocurrence of these violent instabilities and the high associated velocity dispersions required to regulate the disk at a Toomre stability parameter $Q\,\approx\,1$ \citep{DSC09, BE09, Burkert-Elmegreen} and indeed observations are consistent with a dynamical state of the gas close to the $Q\,\approx\,1$ instability limit \citep{genzel11}. Simulations with different numerical schemes, of both idealized galaxies and galaxies in cosmological context, show the rapid development of strong turbulence and giant clump formation in such gas-rich galaxies \citep{noguchi,BEE07,agertz-clumps, ceverino,powell11,martig12, dubois13, hopkins}. 
\medskip	

The violent clump instability thickens the stellar disk \citep{BEM09} and drives an inflow toward the nucleus at a much higher rate than spiral arms and bars \citep{krumholz-flow,bournaud11}. However, other major dynamical effects depend on the life-time of the giant clumps and their ability to survive stellar feedback. Early simulations with no or weak feedback found that the clumps retain much of their mass or increase their mass over time, and survive feedback \citep[][Mandelker et al. in preparation]{immeli,EBE08}. In this case the clumps migrate radially through dynamical friction and gravitational torques in only a few $10^8$\,yr, and convey important amounts of gas and stars inward. This increases the general instability-driven inflow, can rapidly ($\leq\,1$\,Gyr) reshape the disk radial profile into an exponential (Bournaud et al. 2007), form a massive classical, low-Sersic-index bulge \citep{noguchi, EBE08, ceverino}, and increase the nuclear gas flows, powering bright AGN episodes \citep{gabor13, dubois12}. 

However, the ability of giant clumps to survive stellar feedback for $10^8$\,yr or more has been questioned. The momentum conveyed from young stars by photons is an efficient process to affect dense gas structures even before supernovae explode \citep{murray1}, and this has been suggested to rapidly disrupt the giant clumps of high-redshift galaxies where high gas densities could trap the photons \citep{murray2}. A potential issue with short-lived clump scenarios is that observed giant clumps need to live long-enough to accumulate their high baryonic masses. Simulations with strong ad-hoc radiative feedback have shown that giant clumps could form and disrupt within 100\,Myr \citep{genel, hopkins}, but leaving unclear whether rapidly disrupted clumps could reform sufficiently quickly to explain the high observed fraction of clumpy galaxies.

Two observations have raised the possibility of a short clump lifetime. First, outflows have been detected in the vicinity of a few massive clumps, with outflow rates of the order of their star formation rate (SFR) or a few times higher (Genzel et al. 2011, Newman et al. 2012ab), and with high electronic temperatures suggesting strong pressure gradients \citep{lehnert}. Second, the clumps are much fainter in near-infrared imaging \citep{E09}, indicating that their masses are relatively limited in the old stellar components, and indeed the typical ages of the stellar populations contained in clumps are rarely larger than $\leq$\,200\,Myr \citep{Wuyts12}. This indicates that the clumps do not disrupt as rapidly as in simulations with extreme ad-hoc feedback (Genel et al. 2012), and may be rather consistent with typical clump migration timescales of 200--400Myr (Bournaud et al. 2007; Dekel et al 2009b; Ceverino et al. 2010).

On the other hand, it has been proposed that a massive clump wandering in a turbulent disk may rapidly accrete significant amounts of gas \citep{DK13}, which may compensate for the losses in star formation and outflows and make clumps longer-lived. Even if initially warm, gas (re-)accreted this way should become dense enough to cool in less than a dynamical time, hence rapidly refueling the bound star-forming component in the clumps \citep{Pflamm}.

In this paper, we present high-resolution AMR simulations of high-redshift clumpy galaxies using a new, physically-motivated modeling of stellar feedback (Section~2). We show that giant clumps produce outflows and lose aged stars by tidal stripping, but keep their mass and star formation arte about constant over time owing to continuous re-accretion of gas from the surrounding disk (Section~3). The clumps in these models are long-lived and complete their migration inward until they merge with other clumps or with the bulge, and have properties (outflow rates, star formation rates, stellar populations, contrast in optical imaged and stellar mass maps) that appear fully consistent with observations (Section~4).

% - - - - - - - - - - - - - - - - - - - - - - - - - - - - - - - - - - -

\section{Simulations}
\subsection{Models of high-redshift disks}

In the present study, we follow the detailed evolution of giant clumps in high-redshift galaxies, in particular their response to star formation and feedback processes. To this aim we use ``idealized'' (closed-box) simulations of systems with global sizes, masses and gas fractions similar to real galaxies at redshift two, which naturally form giant clumps similar to the observed ones as studied by many previous authors. The advantage of such idealized simulations is that the interstellar medium structure can be resolved with a resolution a decade finer than in state-of-the-art cosmological simulations, resolving the size of heated structures around young star clusters, and sub-structures in the giant clumps through which outflows can propagate. These simulations model disk galaxies with mass distributions, rotation curves and gas fractions that are realistic for $z$\,=\,1-3 galaxies, although their formation history at higher redshifts is not followed. During the evolution of their giant clumps, these simulations lack the replenishment of the disk by external gas infall, which may limit the ability of the clumps to re-accrete gas from the surrounding regions of the disk, thus providing a lower limit to the survivability of clumps\footnote{This effect should not be major because the gas fraction decreases slowly compared to the 400-500\,Myr timescale required for clump migration. In cosmological simulations, the inability to explain the very high gas fractions of $\approx$\,50\% at $z\,\approx\,2$ because of too early consumption into star formation at $z\,>\,3$ may be a larger issue regarding the galactic reservoir of gas in which giant clumps form and evolve (e.g., Ceverino et al. 2010).}. As for dynamical aspects, various studies have shown that both the gravity and turbulence of these galaxies are controlled by internal physics rather than by external infall (e.g., self-gravity: \citealt{Burkert-Elmegreen}, turbulence: \citealt{Hopkins-turb, genel12}). The disk galaxy dynamics can therefore be followed quite reliably in isolation for several hundred Myr.

The fiducial high-redshift disk simulations studied here, labelled G1, G2 and G3 and displayed in Figures~1 to 3, are taken from the {\sc Mirage} sample of disks and mergers presented in Perret et al. (2012a, submitted, arXiv:0768438). An additional model G'2, close to G2 but using a different combination of feedback and resolution parameters, is also used here. 

The initial baryonic masses of galaxies G1 to G3 are 8.7, 3.5 and 1.4 $\times 10^{10}$\,M$_{\odot}$, respectively. Their evolution starts with a gas fraction (gas/gas+stars) of 60\%. The disk scale-lengths are set according to the \citet{vergani} scaling relations, equally compatible with other high-redshift scaling relations \citep[e.g.,][]{Dutton}. The bulge-to-disk stellar mass ratio is 30\%. The star formation model detailed below naturally leads to a realistic specific star formation rate of the order of one Gyr$^{-1}$, realistic for the Main Sequence at redshift $z \approx 1-2$ \citep{ Daddi-MS, Elbaz-MS, nordon}. The evolution is followed for 1\,Gyr, and the average gas fraction over this period is 47\%. Stars and dark matter are modeled with about one million particles each (see details in Perret et al. 2013a).

The alternate model G'2 starts with initial conditions similar to G2 except that the total mass is 17\% higher and the initial gas fraction is 50\%; the main difference is the (weaker) supernovae feedback modeling as detailed below. It also uses a twice higher spatial resolution

%%% In this paper we study the evolution of the main star-forming clumps in these galaxies, the associated local outflows, and their stellar populations. The global evolution of the galactic structure (bulge/disk) and mass (large-scale outflows) will be studied and compared to mergers elsewhere \citep{perret2}. 

\subsection{AMR simulation technique}
We use the Adaptive Mesh Refinement (AMR) code RAMSES \citep{teyssier2002} with a quasi-Lagrangian AMR refinement methodology similar to that used for idealized galaxy simulations in \citet{T10, B10, renaud13}. The grid resolution reaches 7\,pc in the high-density and/or Jeans-unstable regions for our fiducial models G1 to G3. Model G'2 uses a 3.5\,pc resolution . Starting from the coarse level, each AMR cell is refined into $2^3$ new cells if {\em (i)} it contains more than 25 particles, or {\em (ii)} its gas mass content is larger than the ``gas mass resolution'' set equal to $1.5\,\times\,10^4$\,M$_\odot$, or {\em (iii)} the local thermal Jeans length is smaller than four times the current cell size. We also impose a pressure floor to keep the Jeans length greater than four times the smallest cell size and avoid spurious instabilities (\citealt{Truelove}, see Section~2.1 in Teyssier et al. 2010).  
We compute thermal evolution including atomic and fine-structure cooling assuming solar metallicity\footnote{The metallicity of massive galaxies at redshift 1-2 is high enough for their cooling not to be strongly reduced compared to solar metallicity \citep[e.g., ][]{erb06}.}. Such modeling naturally leads to a multiphase ISM, in which structures with number densities up to $10^{5-6}$~cm$^{-3}$ are resolved. The initial gas in the disk has number densities of a few cm$^{-3}$ so that we set the initial temperature to $10^4$~K, which is the typical equilibrium temperature for such moderate densities (e.g., Bournaud et al. 2010, a lower temperature phase appears mostly once denser structures arise). We prevent gas cooling below 100\,K for better computational efficiency\footnote{resolved structures of a few parsecs at 1000~K are dominated by turbulent pressure rather than thermal pressure, so the details of cooling below 1000~K have little net effect at the scales resolved in our models.}.

\subsection{Star formation and feedback}
Star formation is modeled using a constant fraction $\epsilon_{SF} = 1\%$ of gas is converted into stars per local gravitational free-fall time (2\% in the alternative model G'2). This type of local model explains the observed relations between the surface densities of gas and star formation rate in a large range of objects \citep[e.g.][]{Elmegreen02,Renaud12, KDM12}, and when applied to the present simulations matches the gas consumption timescales observed for ``Main Sequence'' galaxies at redshift $z \, \simeq \, 1-2$ (see \citealt{perret1} for the present sample).

Three stellar feedback mechanisms are included:
\begin{itemize}
\item {\em HII region photo-ionization:} around each stellar particle younger than 10~Myr, the photo-ionized region is computed using the Str\"omgren sphere approximation implemented by \citet{renaud13}, the spherical approximation being justified by the fact that each HII region radius is not much larger than the numerical resolution used here (although large non-spherical regions do form from the overlap of smaller spherical regions around clusters of young stellar particles). 
\item {\em Radiation pressure} from young stars is implemented using the \citet{renaud13} model. Once the available momentum in photons is computed for a standard IMF, we use the fact that most of the momentum is carried by ionizing photons, and hence is acquired by the gas that is ionized, i.e. gas that lies in the previously calculated HII regions. This allows us to determine, from physical considerations, how the available momentum $m\times v$ is distributed into a mass $m$ of gas pushed at a velocity $v$: in our model, $m$ is linked to the mass of gas that each stellar particle is able to ionize. This is an important difference compared to other models of high-redshift disks with momentum-driven feedback, where the wind velocity was set by the escape velocity from the clump (or galaxy) to derive the mass loading factor \citep{genel, hopkins}. For the same amount of available momentum, our model is able to generate a higher-velocity wind comprising a lower mass of gas, if high-density gas captures most of the momentum available from young stars. We use a trapping parameter $\kappa=5$ to account for multiple scattering: this is rather high but is realistic as it compensates for other sources of momentum that are not explicitly included, such as (proto)-stellar winds \citep{K12, DK13}.   
\item {\em Supernovae} are implemented following the \citet{dubois08} method, taking into account non-thermal processes as recently proposed by \citet{teyssier13}. To account for non-thermal processes, the dissipation of the injected energy is limited to a a fixed period of 2\,Myr, slower than thermal thermal cooling. This 2\,Myr value is the typical turbulence dissipation timescale expected for the sub-grid structures that our simulations are unable to resolve \citep[see details in ][]{perret1}. This is qualitatively equivalent to the ``delayed cooling'' approach proposed by other authors to model the blastwave phase of supernovae explosions \citep[e.g.,][]{stinson}, but the dissipation timescale is lower here because of the higher resolution.
\end{itemize}

Our alternative model G'2 uses a pure thermal injection for supernovae feedback without a reduced dissipation rate. Hence, the supernova feedback in this model is weaker than in the fiducial runs, while the photo-ionization and radiative pressure feedbacks are the same.

\begin{figure*}
\centering
\includegraphics[width=0.78\textwidth]{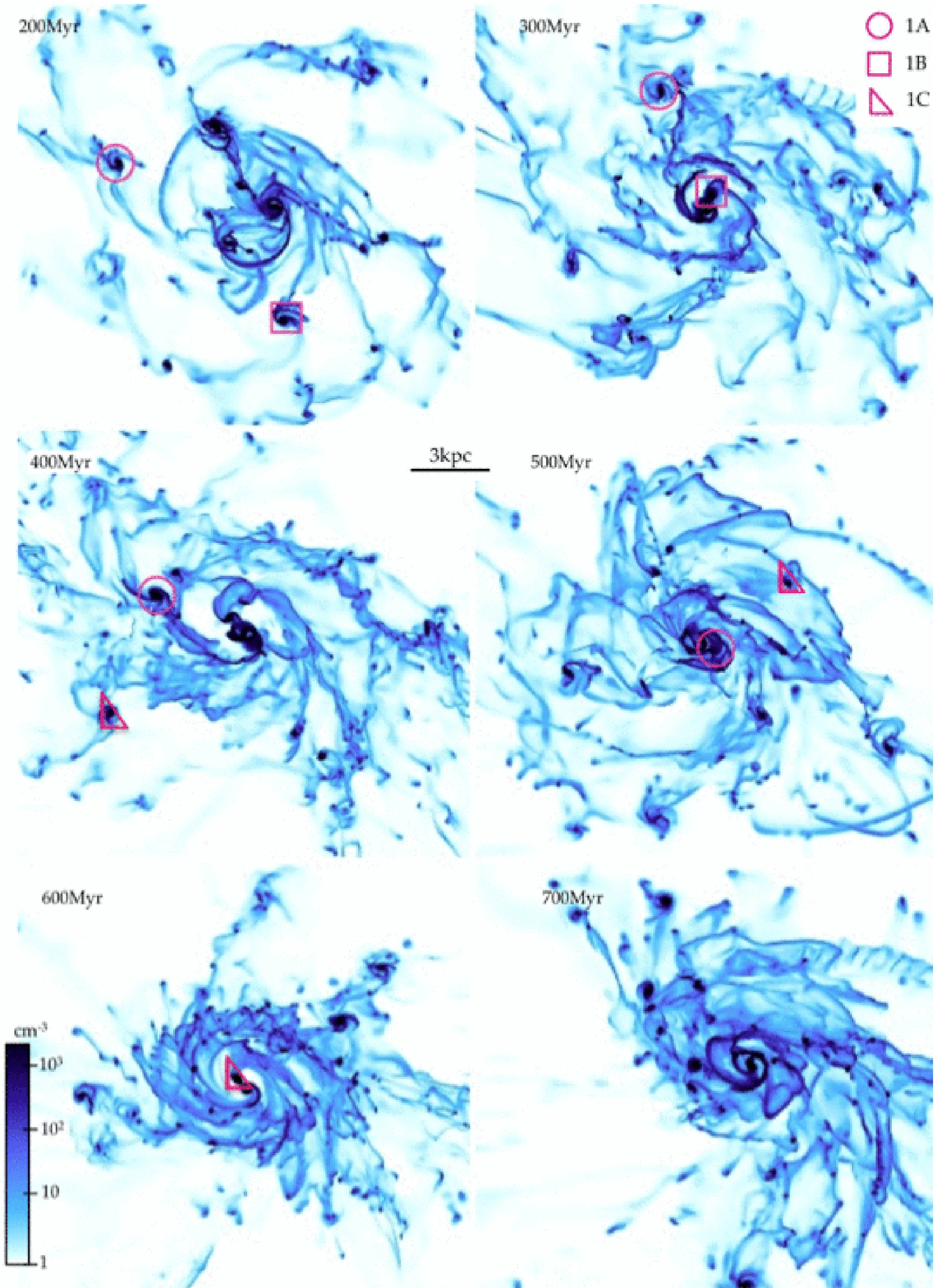}
\caption{\label{galG1} 
Sequence of snapshots showing the mass-weighted average of the gas density along each line-of-sight (i.e., the typical 3-D density reach along each line-of-sight, rather than the column density) for model G1 (high mass). Snapshots are taken every 100\,Myr from 200 to 700\,Myr after the beginning of the simulation. Three typical clumps were selected for analysis, marked with symbols on the maps. At $t$\,=\,700\,Myr, these three typical clumps have merged with the central disk or bulge or with other giant clumps, hence being un-marked in the last snapshost, on which other clumps formed later-on in the outer disk material and/or recycled gas are still present. Detailed sequences and movies of our fiducial models are available in Perret et al. (2013a) and allow the reader to track long-lived clumps in detail.}
\end{figure*}

\begin{figure*}
\centering
\includegraphics[width=0.77\textwidth]{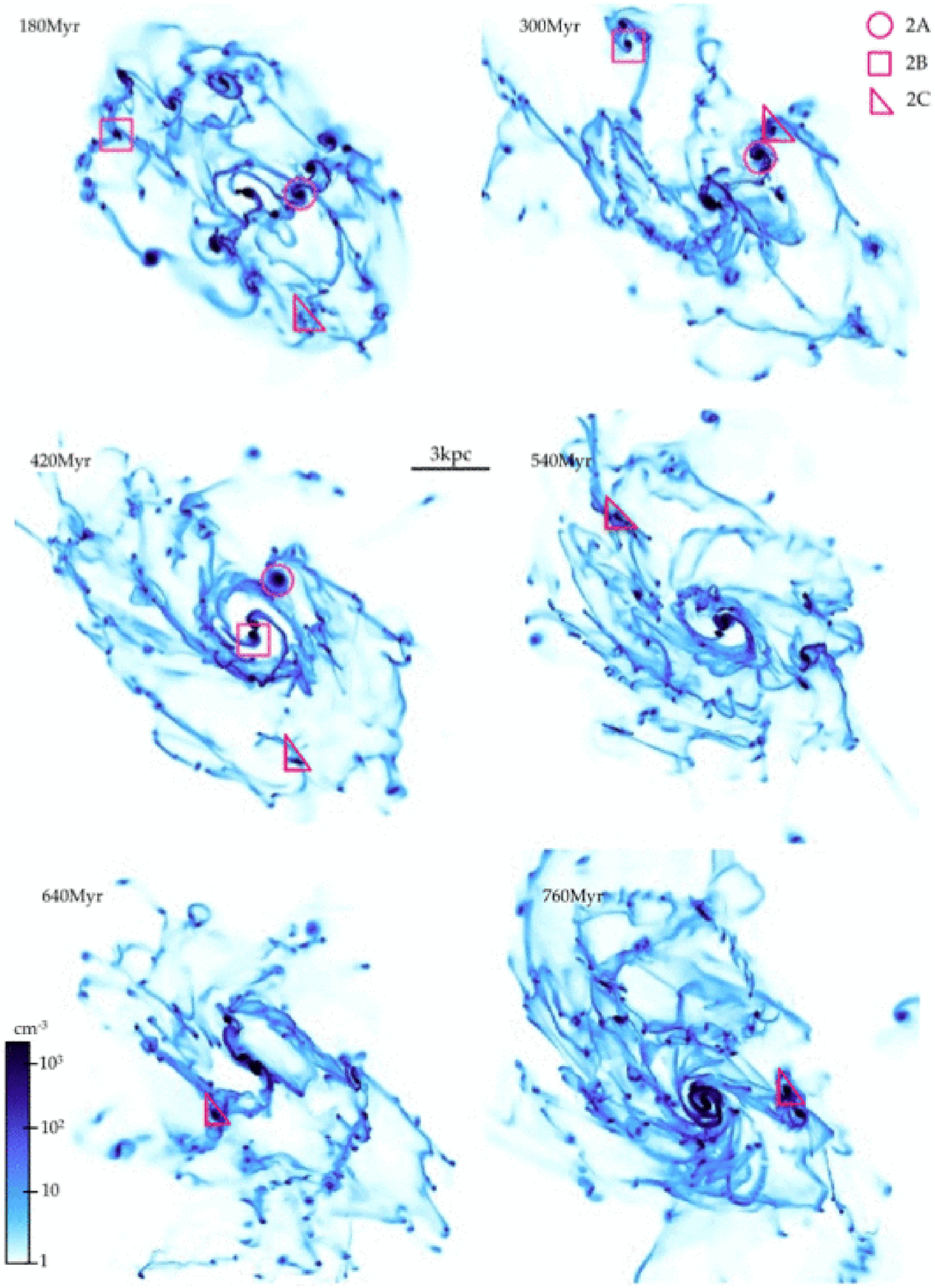}
\caption{\label{galG2} 
Same as Figure~1 for galaxy G2 (medium mass). Detailed sequences and movies of our fiducial models are available in Perret et al. (2013a).}
\end{figure*}

\begin{figure*}
\centering
\includegraphics[width=0.77\textwidth]{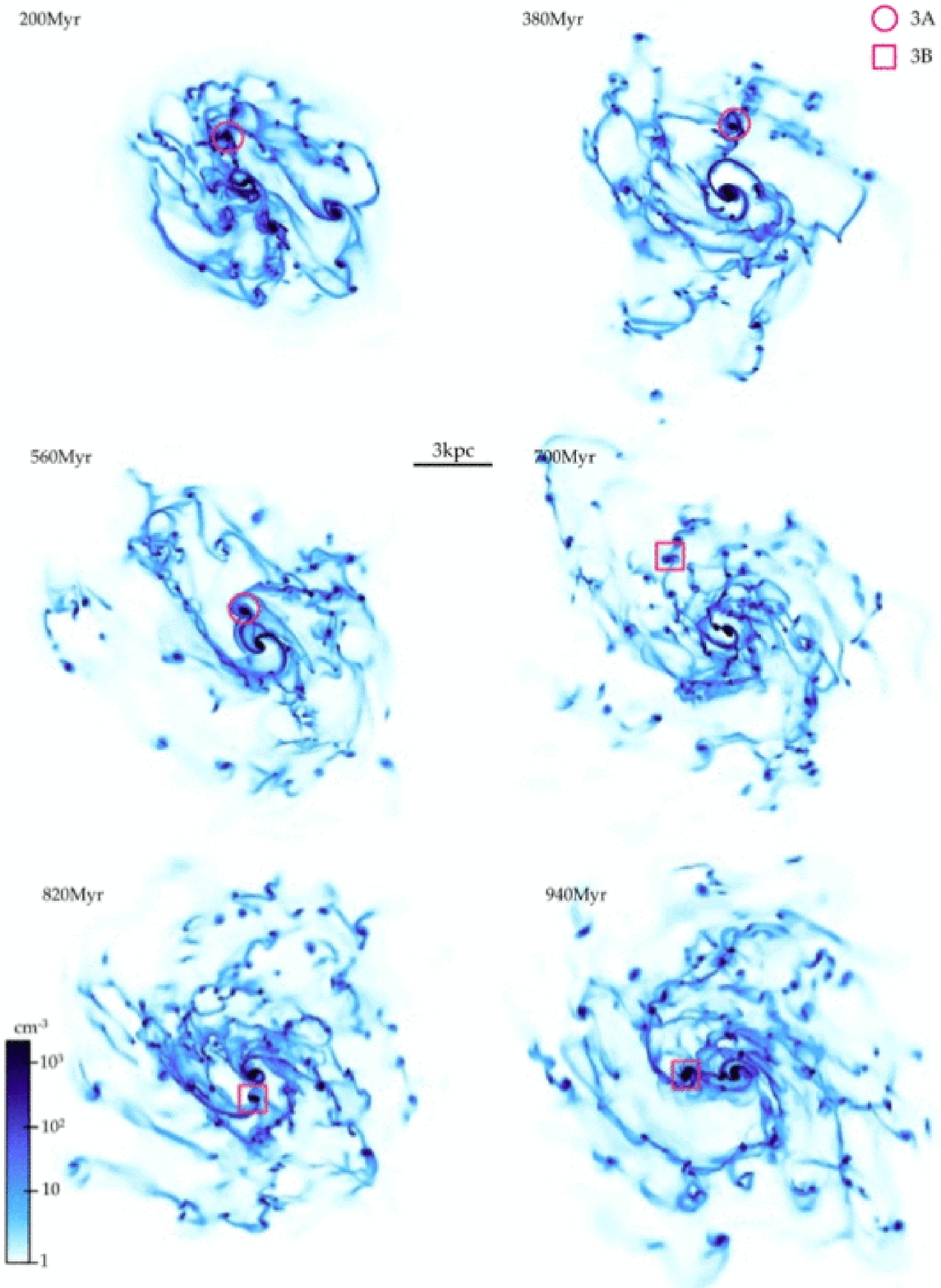}
\caption{\label{galG3} 
Same as Figure~1 for galaxy G3 (low mass). Detailed sequences and movies of our fiducial models are available in Perret et al. (2013a).}
\end{figure*}

\subsection{Selection and measurement of giant clumps}
The giant clumps formed in the simulations are tracked visually with an output frequency of 4\,Myr, and are analyzed in detail every 40\,Myr. We do not consider here the clumps that undergo mergers with other clumps of similar mass, to make sure that a main clump progenitor can be unambiguously identified. 

For each snapshot the visually determined position of each clump is used as an initial guess for its mass center. We compute the mass center of the baryons included in a sphere of radius 200\,pc around this initial guess, and iterate the procedure around the newly determined position, until the position shifts by a distance smaller than the resolution of the simulations. Once the mass center of the clump is found this way, we determine the clump radius and mass with the following procedure. Using a radius $R_1$ initially set to 100\,pc, we compute the mass in the sphere of radius $R_1$, and check the mass in a sphere with twice large volume (i.e., radius $2^{1/3}\, R_1$). If the mass increase is larger than 30\%, we consider that the clump material was not mostly encompassed by the initial boundary, and iterate the process with an initial radius $R_1+100$\,pc, until the mass converges. Convergence is never reached at 100\,pc, and always reached at radius $\leq \,$ 500\,pc, namely the clump radii range from 200\,pc to 500\,pc. These positions and radii are used for all the following measurements: gaseous and stellar masses, mass flows, star formation rate per clumps, etc. Given the method used, clump masses are determined with an uncertainty better than 30\%, the uncertainty on clump radii can be up to 100\,pc but only the masses are relevant in the following study.

% - - - - - - - - - - - - - - - - - - - - - - - - - - - - - - - - - - -

\begin{figure*}
\centering
\includegraphics[width=0.99\textwidth]{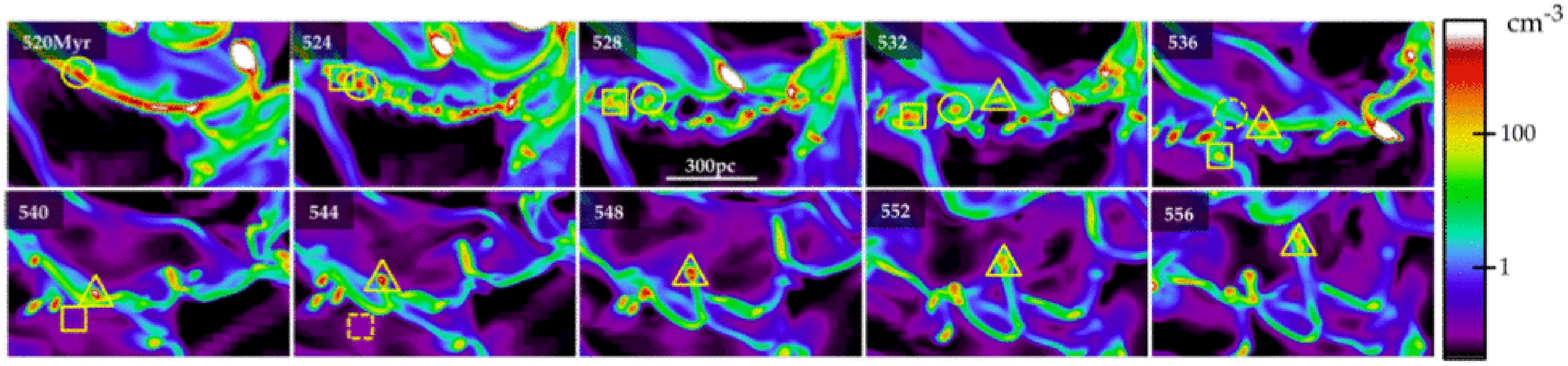} %small_clouds.eps}
\caption{\label{small_clouds} Sequence detailing the continuous formation and rapid dissolution of low-mass clouds, at the middle of the time evolution of model G2. The snapshot size is 0.8x1.6~kpc, and time is indicated in Myr, with one snapshot every 4\,Myr. A moderately dense spiral arm, free of giant clump, forms a first generation of clouds. Two of these clouds are identified with the circular and boxy shapes: they leave only weak and dissolving knots after 15--20\,Myr. Other clouds form at different locations, such as the one first indicated with the triangle at $t$\,=\,532, which again dissolves, leaving a lower-density knot seen at $t$\,=\,556 and unseen at $t$\,=\,600\,Myr.}
\end{figure*}

\begin{figure*}
\centering
\includegraphics[width=0.87\textwidth]{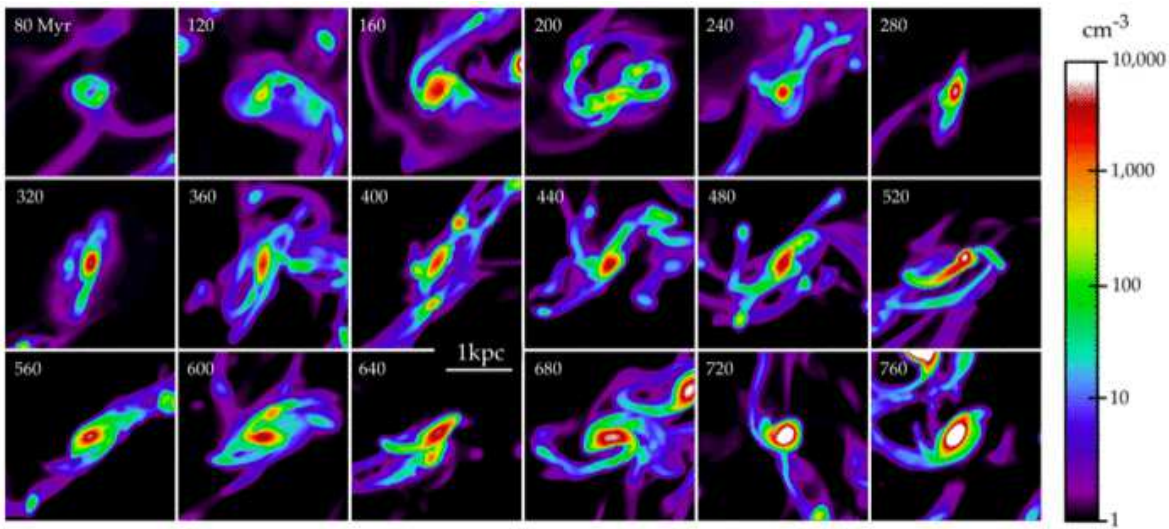}
\caption{\label{seq_2C} 
Zoomed views of gas in the long-lived clump 2C, with an average baryonic mass of $8\times 10^8$\,M$_\odot$. The snapshots show the mass-weighted average gas density, with one snapshot every 40\,Myr. Between the third and fourth panels ($t$=160-200Myr), the clump accretes another clump (about half its mass), which triggers an increase in its SFR, and a later increase in the local outflow rate (see Fig.~\ref{rate_curves}); the clump gets a more disturbed appearance but the baryonic potential well in place rapidly re-accretes gas and the clump survives this local enhancement of the stellar feedback. Another such event, triggered by the accretion of surrounding diffuse gas and small clouds, occurs between the seventh and eighth panels ($t$\,=\,320-360\,Myr).}
\end{figure*}

\begin{figure}
\centering
\includegraphics[width=0.45\textwidth]{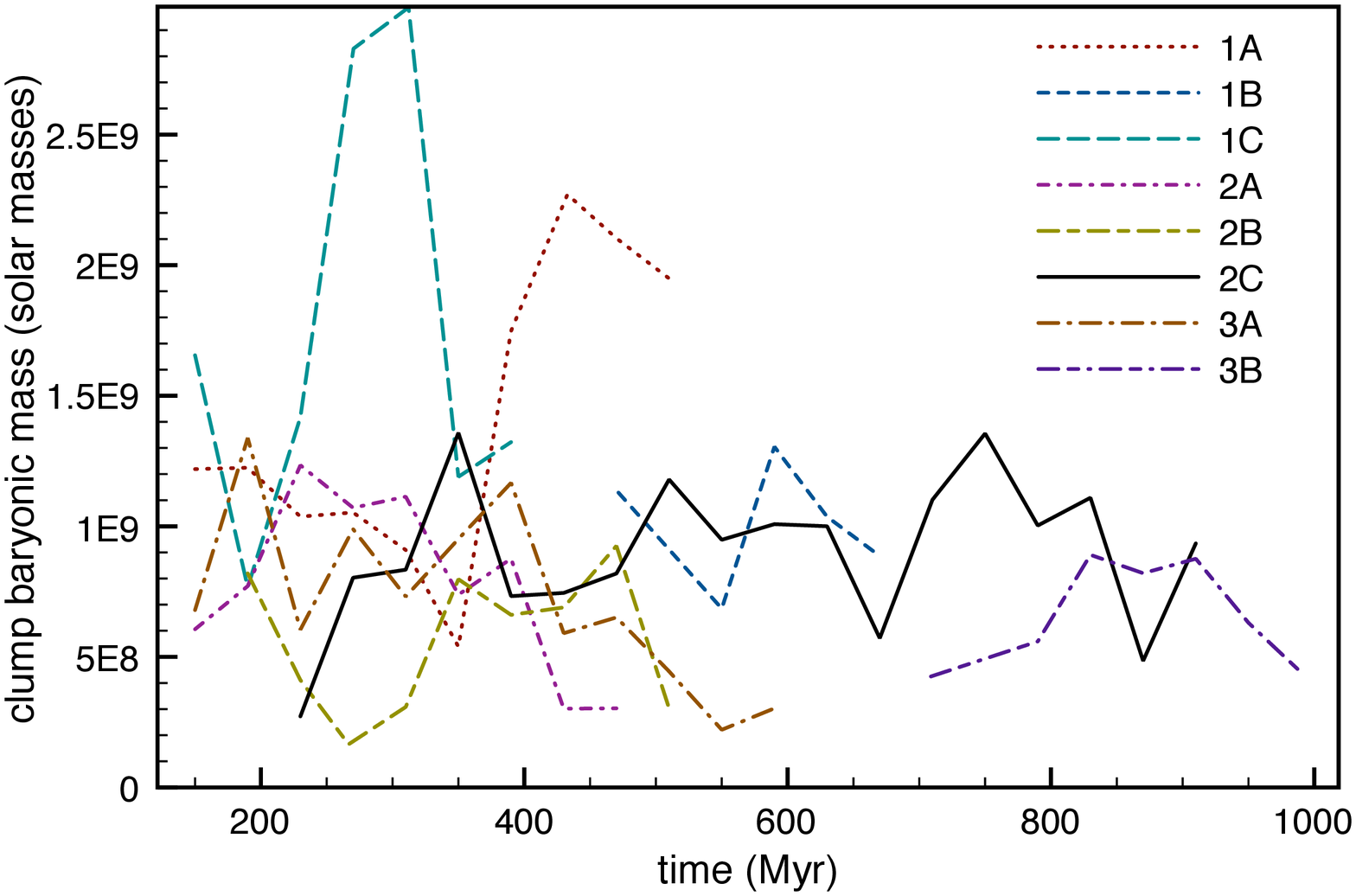}
\caption{\label{M_t_clumps} 
Evolution of the baryonic (gas+stars) mass of clumps as a function of time for the clumps tracked in the simulations.
}
\end{figure}

\section{Results}
\subsection{Short-lived gas clouds and long-lived giant clumps}

In our simulations, gas clouds below a mass of a few $10^7 M_\odot$ are short-lived. This is also true for gas clouds in low-redshift galaxies with the same technique (e.g. Renaud et al. 2013). A detailed time sequence showing the rapid formation and dissolution of low-mass gas clouds (not giant clumps) in our current model G2 is shown on Figure~\ref{small_clouds}. Note that simulated molecular clouds in low-redshift galaxies were found to be short-lived and un-virialized even when only some of the feedback modes are included \citep[e.g.,][]{tasker,B10,ceverino-2}.

In contrast, the giant clumps more massive than $\approx$\,$10^8$\,M$_\odot$ remain long-lived in our simulations (Fig. 1 to 3) . Such clumps form by gravitational instability in high-redshift disks, with high Jeans length and mass due to the high level of turbulence \footnote{regardless of the exact source of turbulent energy, gravity and star formation are the main internal energy sources and saturate at a $Q\approx 1$ level.} required to self-regulate the disk at a Toomre parameter $Q \approx 1$. Indeed the average value of the gas velocity dispersion\footnote{one-dimensional r.m.s. dispersion measured in boxes of (200\,pc)$^3$} at the disk scale-length and after two rotation periods ranges from 38 km\,s$^{-1}$ (G3) to 53 km\,s$^{-1}$ (G1). This process of giant clump formation in a turbulent medium has been studied in many works and is not detailed again here \citep{noguchi, immeli, BEE07, agertz-clumps, DSC09, ceverino, genel, hopkins}.

These giant clumps are long-lived and persist until coalescence with the central bulge or with another bigger clump. Most of them can be tracked in the simulations for 200--500\,Myr and sometimes up to 700\,Myr. We display in Figure~\ref{seq_2C} a time series showing the detailed evolution of a very long-lived clumps in our sample (clump 2C from model G2 displayed in Fig.~2). The reason why this clump is particularly long-lived (at least 700\,Myr) is that it forms at a large radius and keeps a low-eccentricity orbit, maximizing the timescale for inward migration by dynamical friction, while most giant clumps would reach the central bulge within 500\,Myr\footnote{for galaxies in the mass range studied here.}. As will appear in the following parts, this clump does not have extreme properties in terms of internal mass, size, or formation rate, and is simply used to illustrate the evolution of its mass content on a timescale that is not limited by rapid central coalescence, as would be the case with some clumps formed on different radii and/or on different orbits.

Over their long lifetime, the mass of the giant clumps remains relatively constant, without a major increase or decrease over time, in spite of the gas outflows detailed hereafter. We show in Figure~\ref{M_t_clumps} the time evolution of the baryonic mass of each of the clumps tracked in detail in our models: these masses, ranging from a few $10^8$ to $2\times 10^9$\,M$_\odot$ and occasionally a bit higher, and fluctuate about roughly constant values, without any significant global mass increase or decrease. As we detail in the following sections, giant clumps lose mass through high-velocity gas outflows and dynamical escape of aged stellar populations, but also accrete gas from the surrounding disk. The rate of mass accretion in our simulations (detailed in Sect.~3.4) compensates for the losses, and is consistent with the theoretical estimates from \citet{DK13} or \citet{Pflamm}. Hence realistic feedback does not disrupt the giant clumps on timescales shorter than $10^8$~yr, even when radiative pressure and non-thermal effects in supernova bursts are included, and continuous capture of baryons maintains their mass high on the long term.

\subsection{Gas outflows from massive clumps}

The time evolution of the clump masses (Fig.~\ref{M_t_clumps}) shows that, in spite of being on average roughly constant, their baryonic mass can sometimes decrease. The detailed sequence of clump 2C (Fig.~\ref{seq_2C}) also shows that the mass and density of the gas in the clumps can sometimes be significantly reduced. While the reduction of the gas mass alone could be the result of consumption by star formation, the fact that the total baryonic mass can sometime decrease implies that gas can actually be lost by the clumps, not just converted into stars.

\begin{figure*}
\centering
$\,$ \vspace{.2cm}\\
\includegraphics[width=0.68\textwidth]{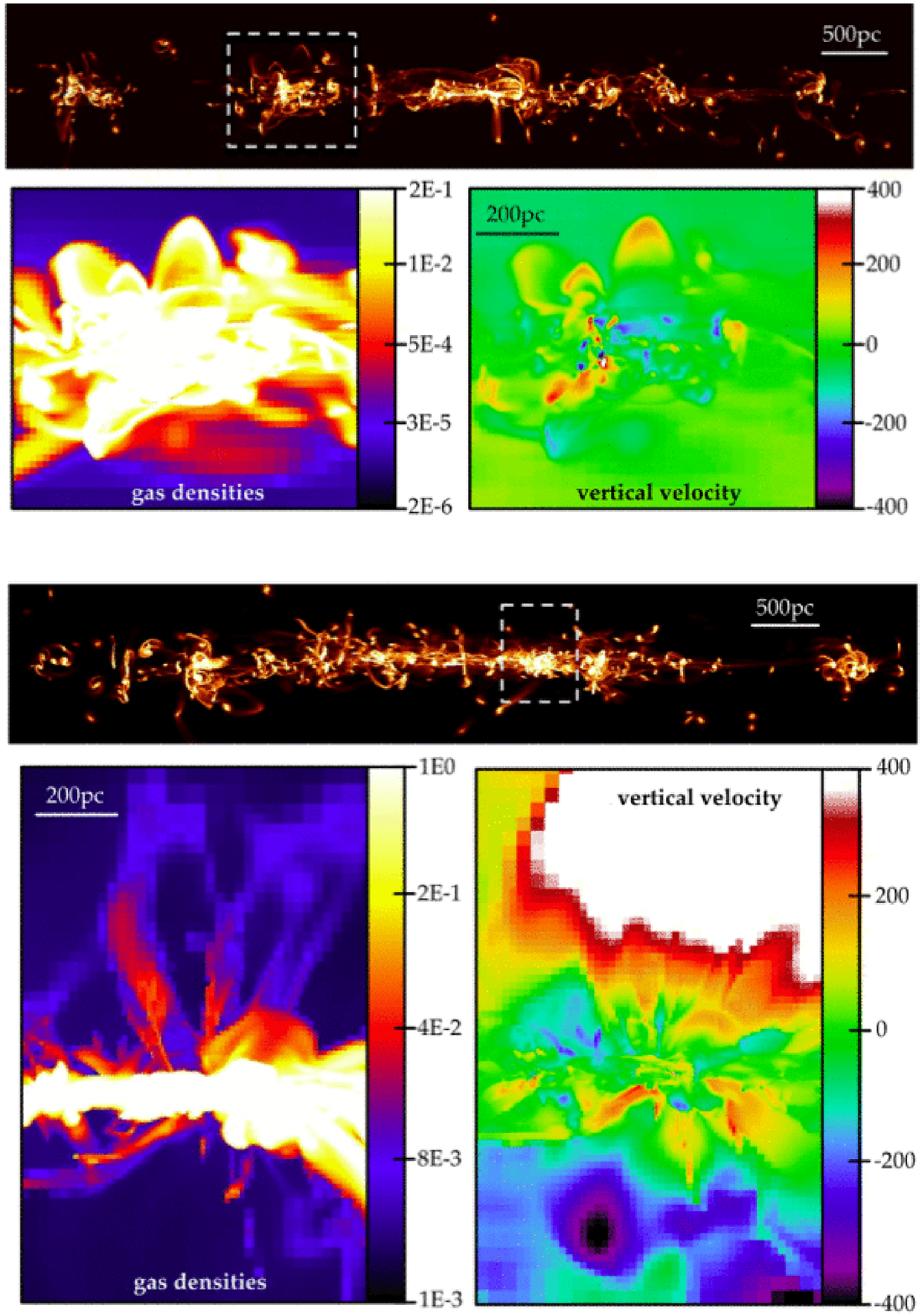} 
\caption{\label{flow_maps} 
Birth of a gas outflow from a long-lived giant clump in model G'2. The entire galaxy is shown edge-on to the top (gas column density maps) after 90 and 270\,Myr in the simulation, i.e. shortly after clump formation and at a more evolve stage. A giant clump is tracked and is shown in the inset maps of the mass-weighted average number density of gas (in cm$^{-3}$) and vertical velocity perpendicular to the disk plane (in km\,s$^{-1}$). At the early stage, shock fronts are formed by the outflowing gas onto the hot and diffuse halo surrounding the galaxy, with number densities of $10^{-2}\,-\,10^{-1}$\,cm$^{-3}$ and velocities around 200\,km\,s$^{-1}$. Later-on a higher-velocity outflow escapes the clump region at velocities of 300--400\,km\,s$^{-1}$ with densities of a few $10^{-2}$\,cm$^{-3}$, corresponding to an outflow rate of 1--2\,M$_\odot$\,yr$^{-1}$ across a section of 1\,kpc$^2$. Note that the outflow continues to be accelerated at more than 1\,kpc above the disk mid-plane, because of a pressure gradient, which turns the local outflow into a global galactic-scale outflow. A movie is available ({\tt http://youtu.be/Qm5-SkgnDYs}) to show the development and expansion of gaseous outflows from clumps to galactic scales in this run.}
\end{figure*}

The giant clumps in our simulations produce gaseous outflows. The birth and evolution of high-velocity outflows from the clumps is illustrated in Figure~\ref{flow_maps}, where the same clump is shown after 90\,Myr (when it just formed) and 270\,Myr of evolution. At the early stage, shock fronts are formed by the outflowing gas onto the hot and diffuse halo surrounding the galaxy, with number densities of $10^{-2}\,-\,10^{-1}$\,cm$^{-3}$ and velocities around 200\,km\,s$^{-1}$. Later-on a higher-velocity outflow escapes the clump region at velocities of 300--400\,km\,s$^{-1}$ with densities of a few $10^{-2}$\,cm$^{-3}$, corresponding to an outflow rate of 1--2\,M$_\odot$\,yr$^{-1}$ across a section of 1\,kpc$^2$. It is interesting to note that the gaseous outflow (1) can largely exceed the local escape velocity and (2) continues to be accelerated more than 1\,kpc above the disk mid-plane, because of a pressure gradient that develops in the stead-state outflow regime. This turns into a galactic-scale outflow, the properties (rate, velocity) of which will be analyzed elsewhere (Perret et al. 2013b). A movie is available\footnote{{\tt http://youtu.be/Qm5-SkgnDYs} } and shows the formation of giant clumps followed by the development and expansion of gaseous outflows from clumps to galactic scales in model G'2.

\begin{figure}
\centering
\includegraphics[width=0.49\textwidth]{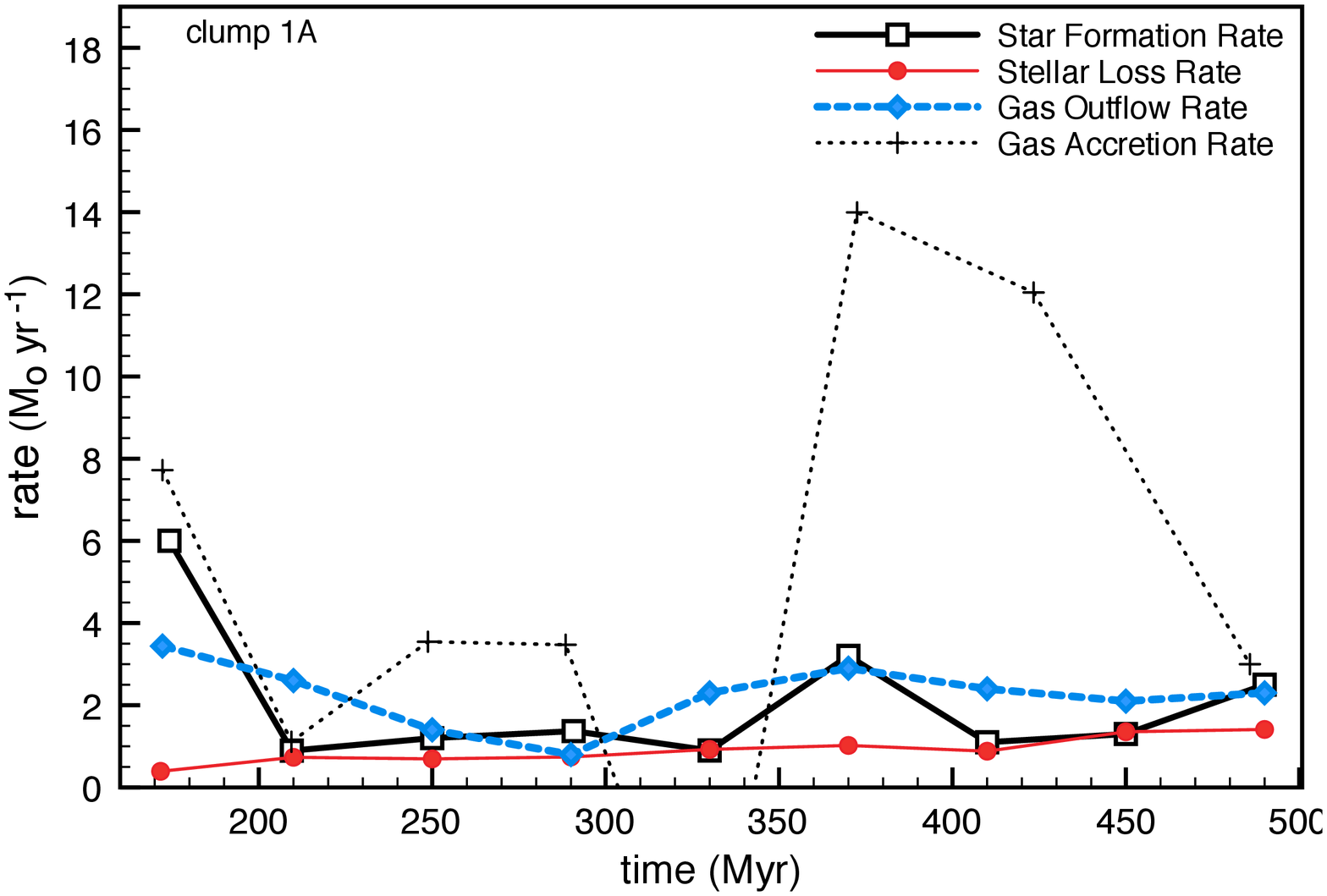}
\includegraphics[width=0.49\textwidth]{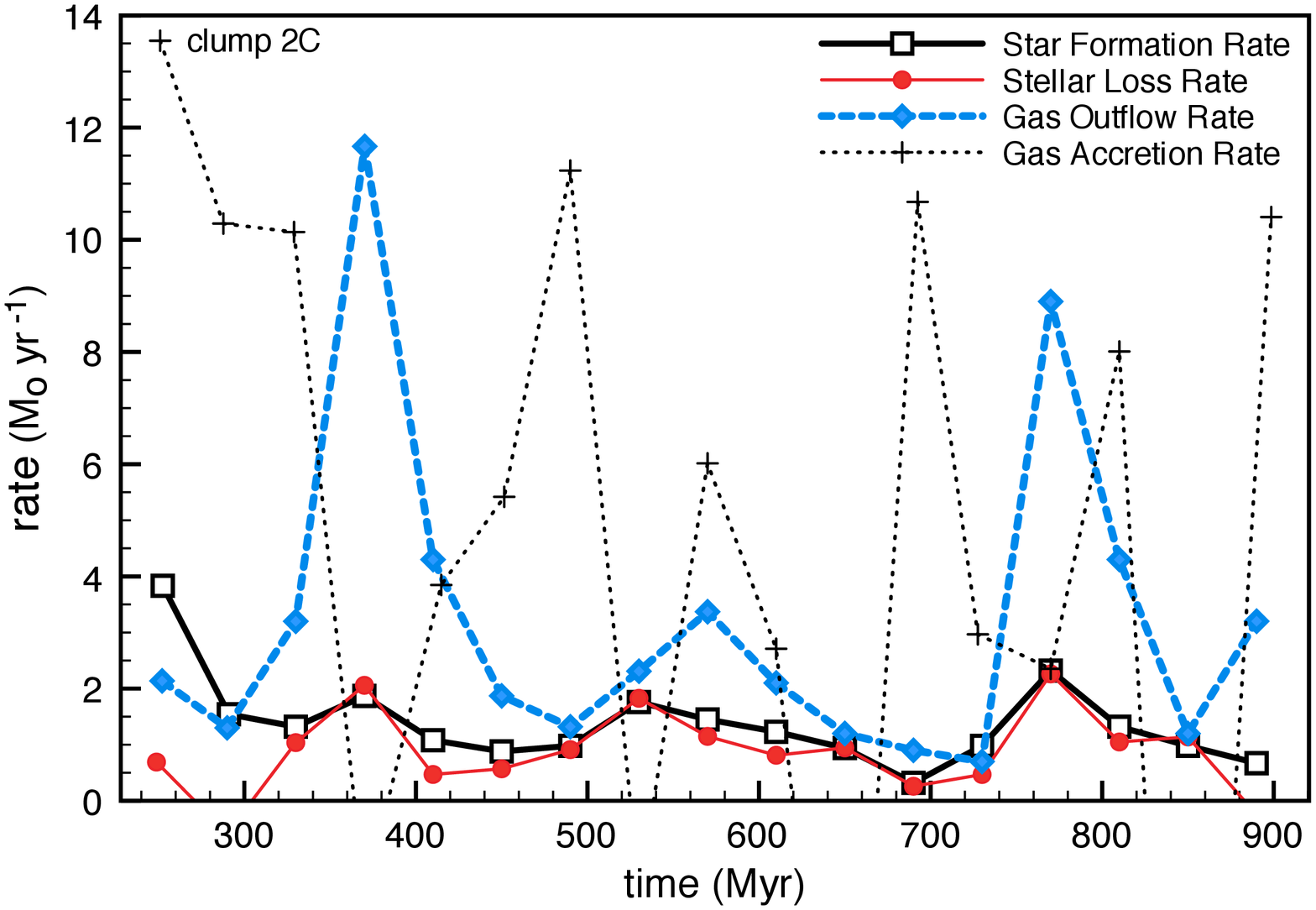}
\includegraphics[width=0.49\textwidth]{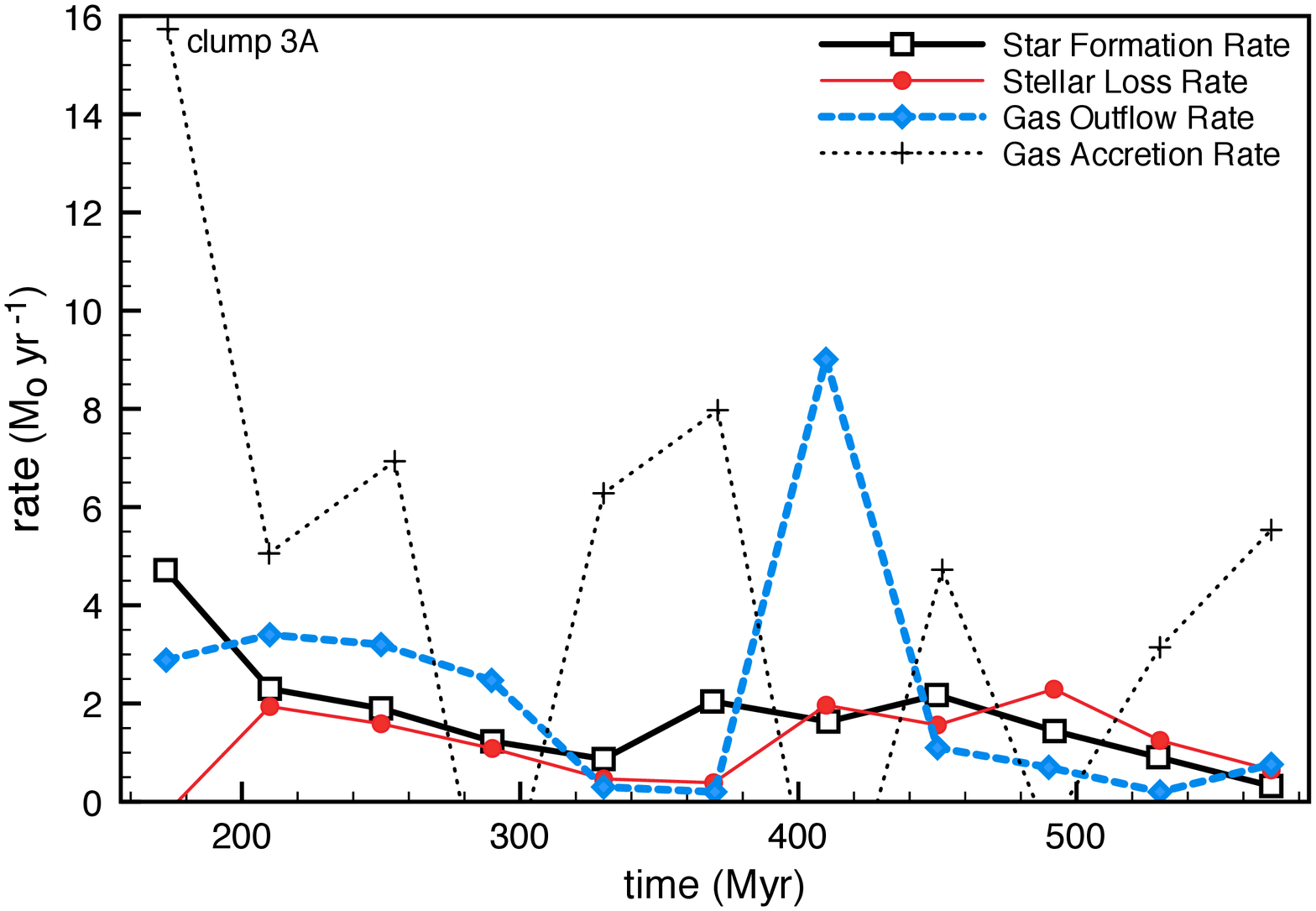}
\caption{\label{rate_curves} Time evolution of the star formation rate (thick solid line), outflow rate (thick dashed line), gas infall rate (thin dashed line), stellar evaporation rate (thin solid line), for three representative clumps. We picked the longest-lived clump in each of the three fiducial simulations to increase the sampling of possible events. Gaussian smoothing of FWHM 40\,Myr was applied to all quantities for clarity. The star formation rate and outflow rate typically increase after periods of intense accretion of the surrounding diffuse gas and/or smaller clumps. There is also an initial burst of star formation when the each clump initially collapses ($t$=0 is the beginning of each simulation, and the first data points are 20\,Myr after the first detection of each clump).}
\end{figure}

\begin{figure}
\centering
\includegraphics[width=0.4\textwidth]{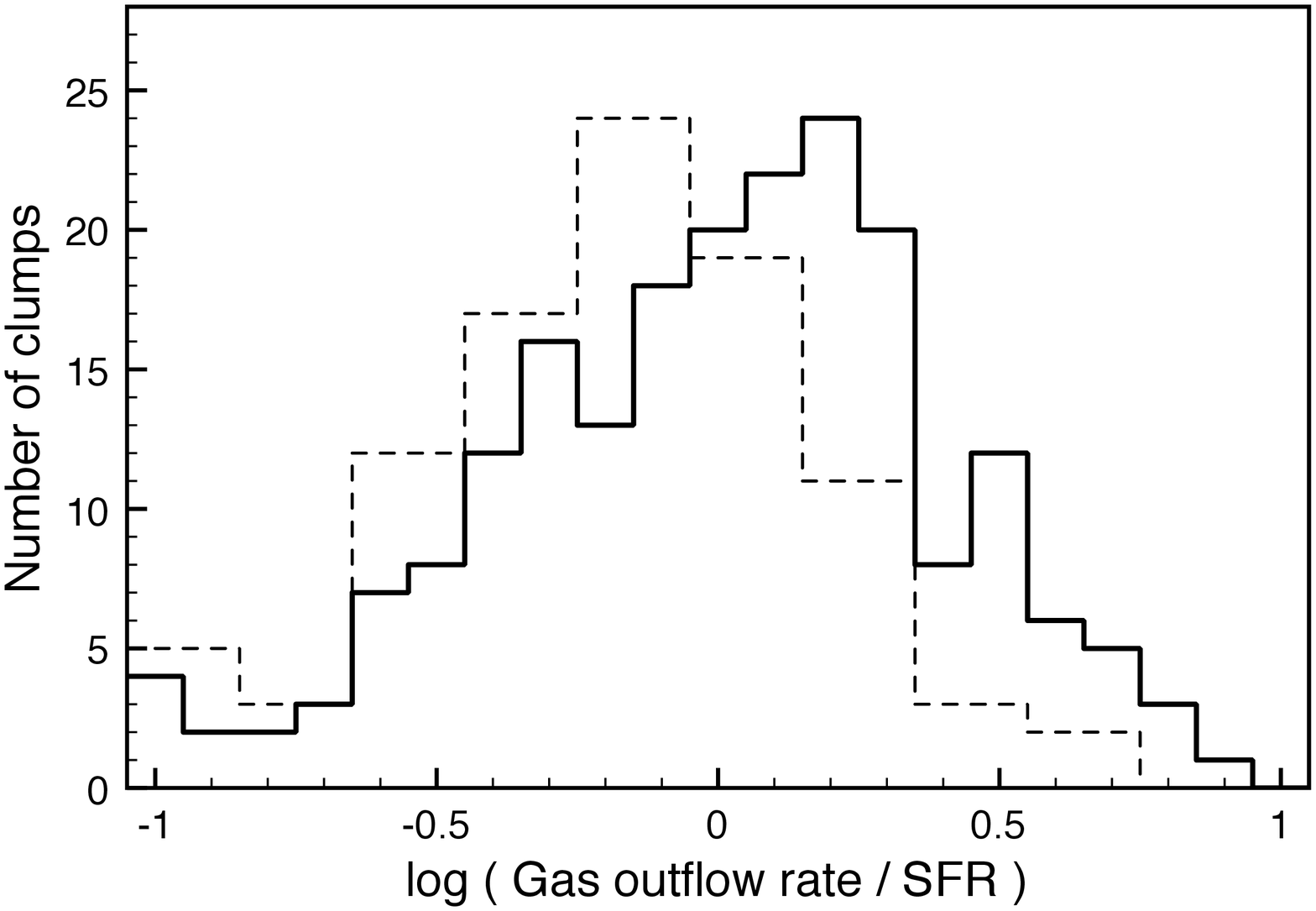}
\includegraphics[width=0.4\textwidth]{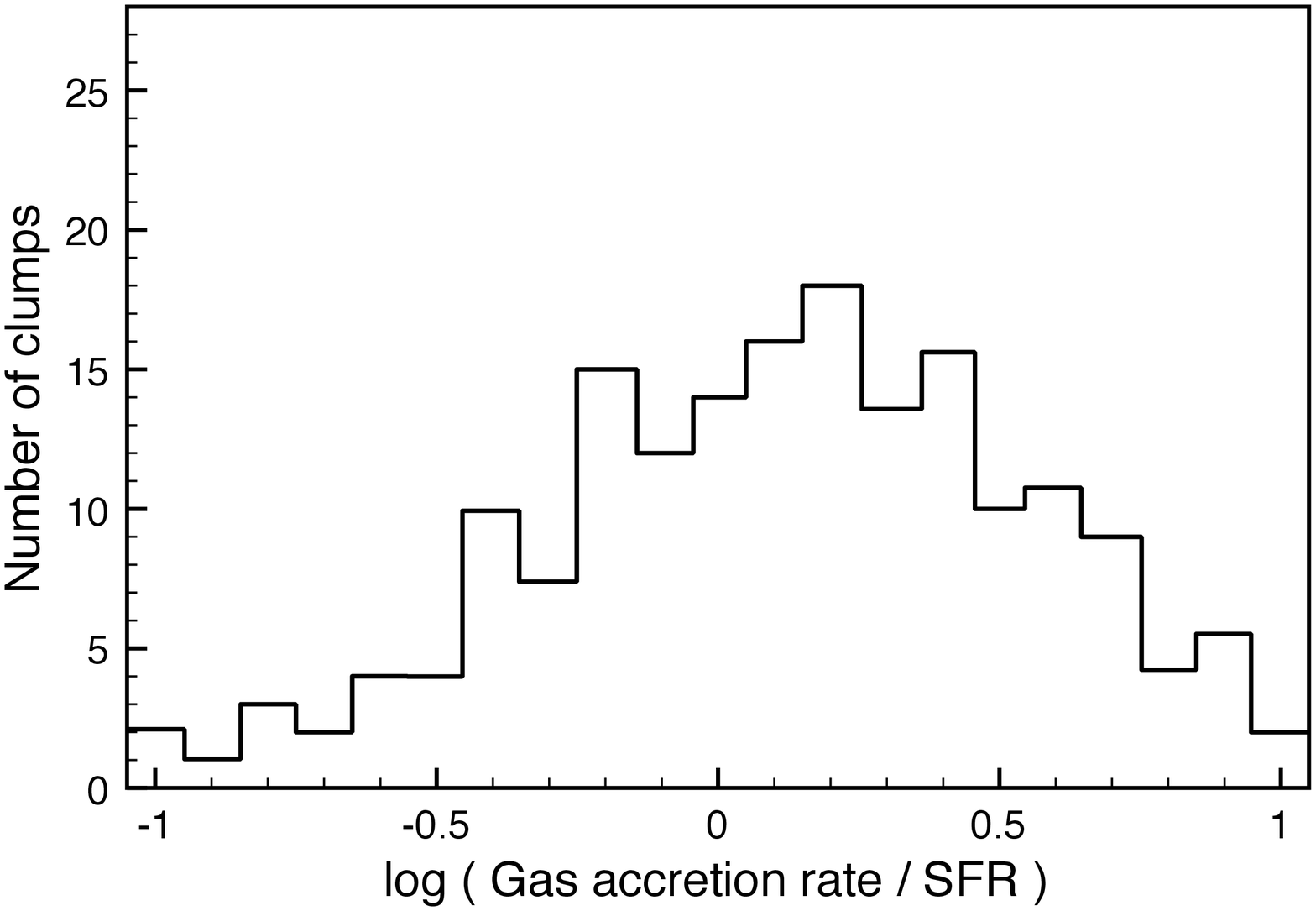}
\caption{\label{out_sfr_ratio} Top: Statistical distribution of the outflow rate to SFR ratio at the scale of giant clumps. The thick histogram is for the cumulated distribution in the fiducial runs G1, G2 and G3. The dashed histogram is for run G'2 (normalized to the same maximum), in this run weaker supernovae feedback is employed along with the same radiation pressure model. The difference shows that the birth of outflows is not ensured solely by radiation pressure, but that supernovae explosions or at least their coupling with radiation pressure play a significant role. Bottom: rate of gas accretion by the clumps, for the fiducial runs G1, G2 and G3. It is of the order or slightly larger than the gas outflow rate.}
\end{figure}

We have systematically measured the outflow rate from each clump in two different ways: through a spherical boundary around the clump (identical to the boundary used to measure the clump mass), and through a pair of 1\,kpc$\,\times\,$1\,kpc planar boundaries positioned\footnote {parallel to the initial disk plane} 1\,kpc above and below the disk mid-plane. In the following we use the latter measurement, after noticing that the former definition leads to similar measurements of the outflow rates from clumps with some additional uncertainty\footnote{Measurements across the spherical boundary may include diffuse gas that passes next the clump and enters/leaves the boundary. Hence the choice of the planar boundaries above and below the disk plan ensures to capture the feedback-induced outflow.} (on average 23\% higher, with an r.m.s. relative deviation between the two measurements of 37\%) . The fact that these measurements yield similar results indicates that the outflows are significantly non-isotropic even at only 1\,kpc from the clump centers, as can be seen in the examples of outflow velocity fields displayed in Figure~\ref{flow_maps}.

The typical outflow rate from the giant clumps in our simulations is 1--4\,M$_\odot$\,yr$^{-1}$, and up to 12\,M$_\odot$\,yr$^{-1}$ during short episodes, of the order of the star formation rate in each clump and up to a few times higher. The time evolution of the outflow rate is shown for three individual clumps on Figure~\ref{rate_curves}. Figure~\ref{out_sfr_ratio} shows the statistical distribution of the outflow rate to star formation rate ratio for our main sample and for simulation G'2 using weaker supernovae feedback (and higher resolution). 

\bigskip

\begin{figure}
\centering
\includegraphics[width=0.45\textwidth]{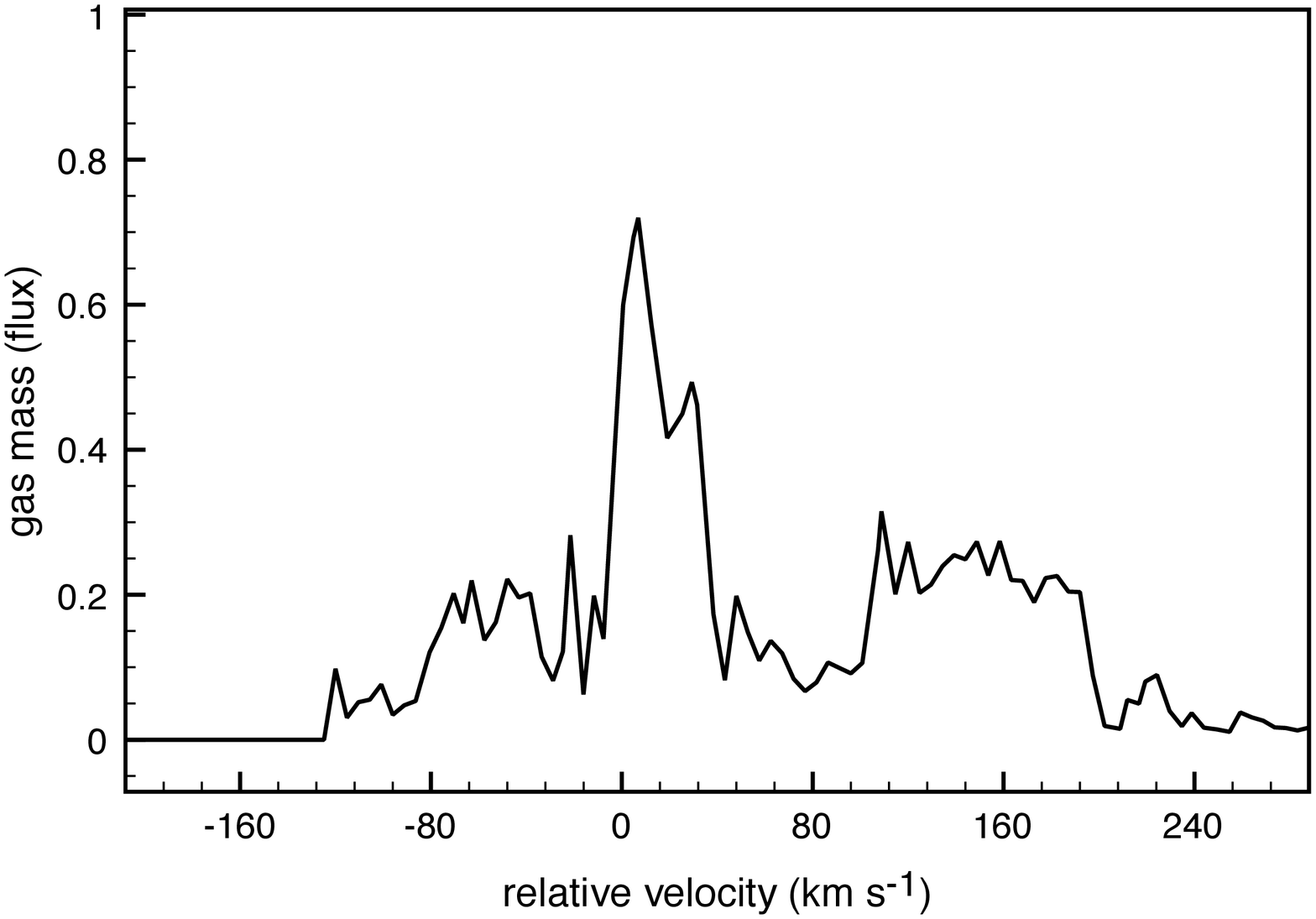}
\includegraphics[width=0.45\textwidth]{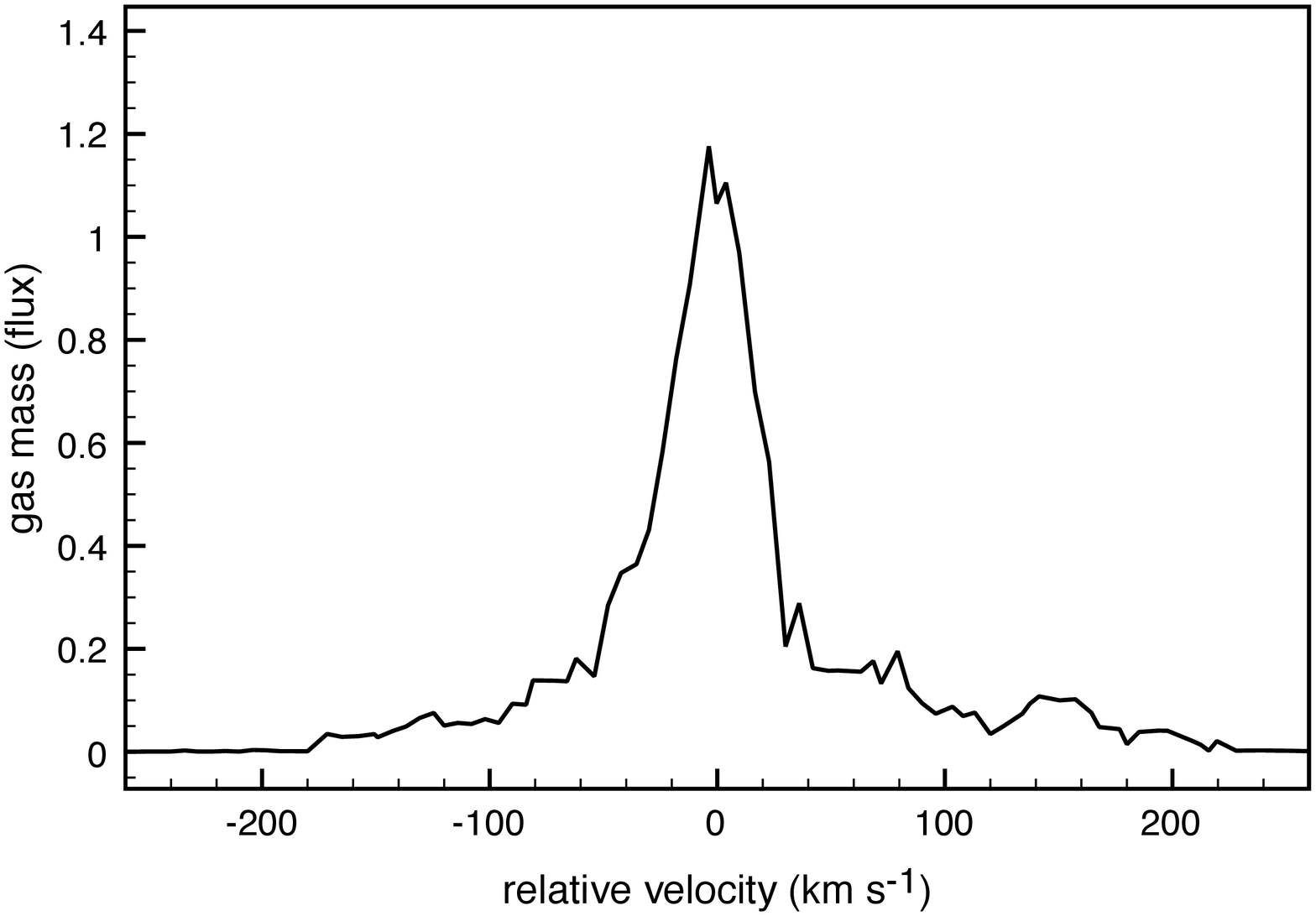}
\caption{\label{spec_2C} 
Line-of-sight velocity distribution (i.e., synthetic spectrum) of clump 2C, observed with a beam 600~pc FWHM, with an almost face-on orientation of the host galaxy. Top: the system is shown during its peak of outflow activity after a big gas cloud was absorbed (see Fig.~\ref{seq_2C} and \ref{rate_curves}), the clump spectrum is broadly irregular with several high-velocity components. Bottom: we show the system in a calmer phase, 80\,Myr later,  when the mass outflow rate is marginally higher than the SFR. A double-Gaussian profile is observed, as in the stacked spectrum for all clumps shown in Figure~\ref{spec_stack}.}
\end{figure}

\begin{figure}
\centering
\includegraphics[width=0.4\textwidth]{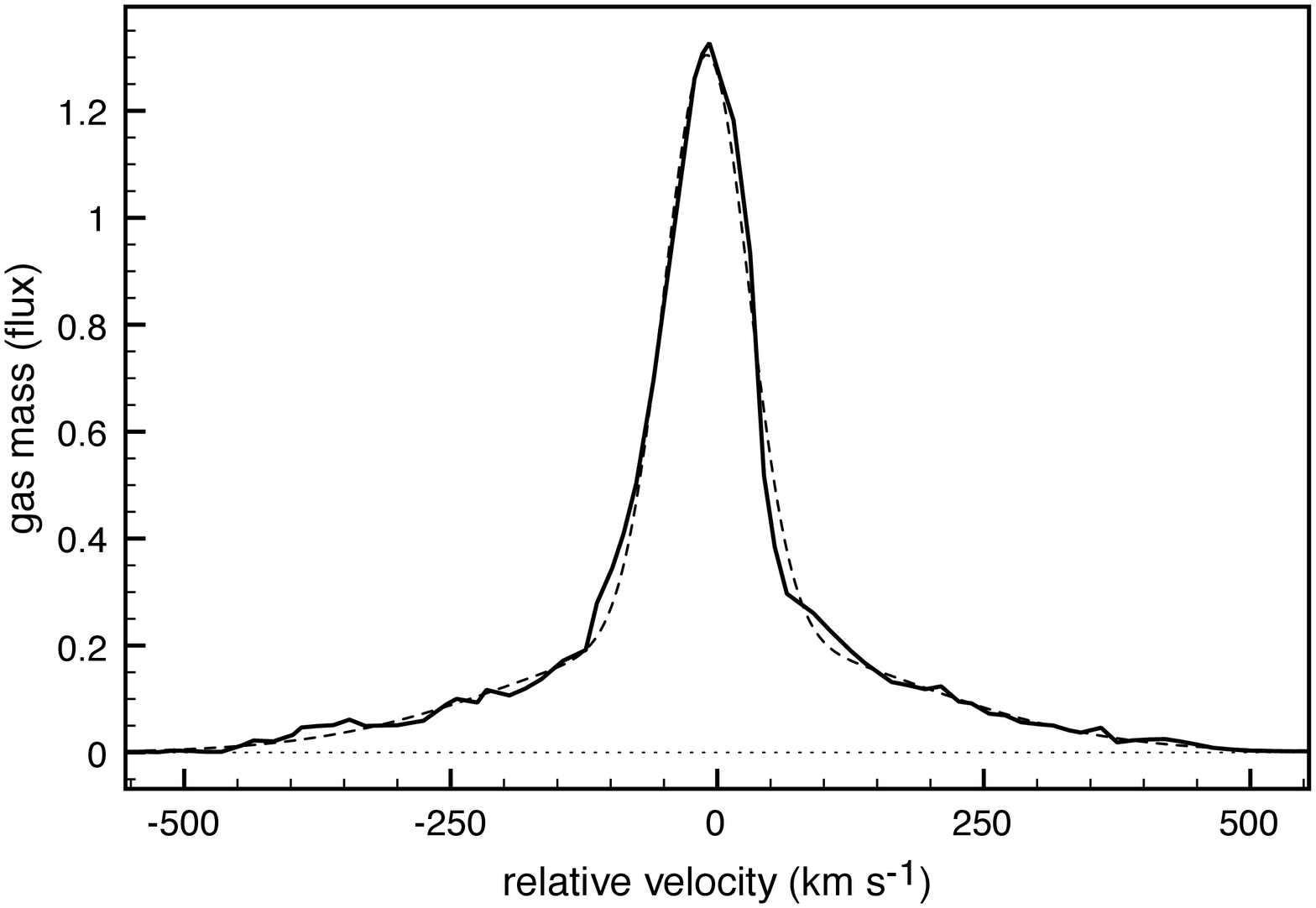}
\includegraphics[width=0.4\textwidth]{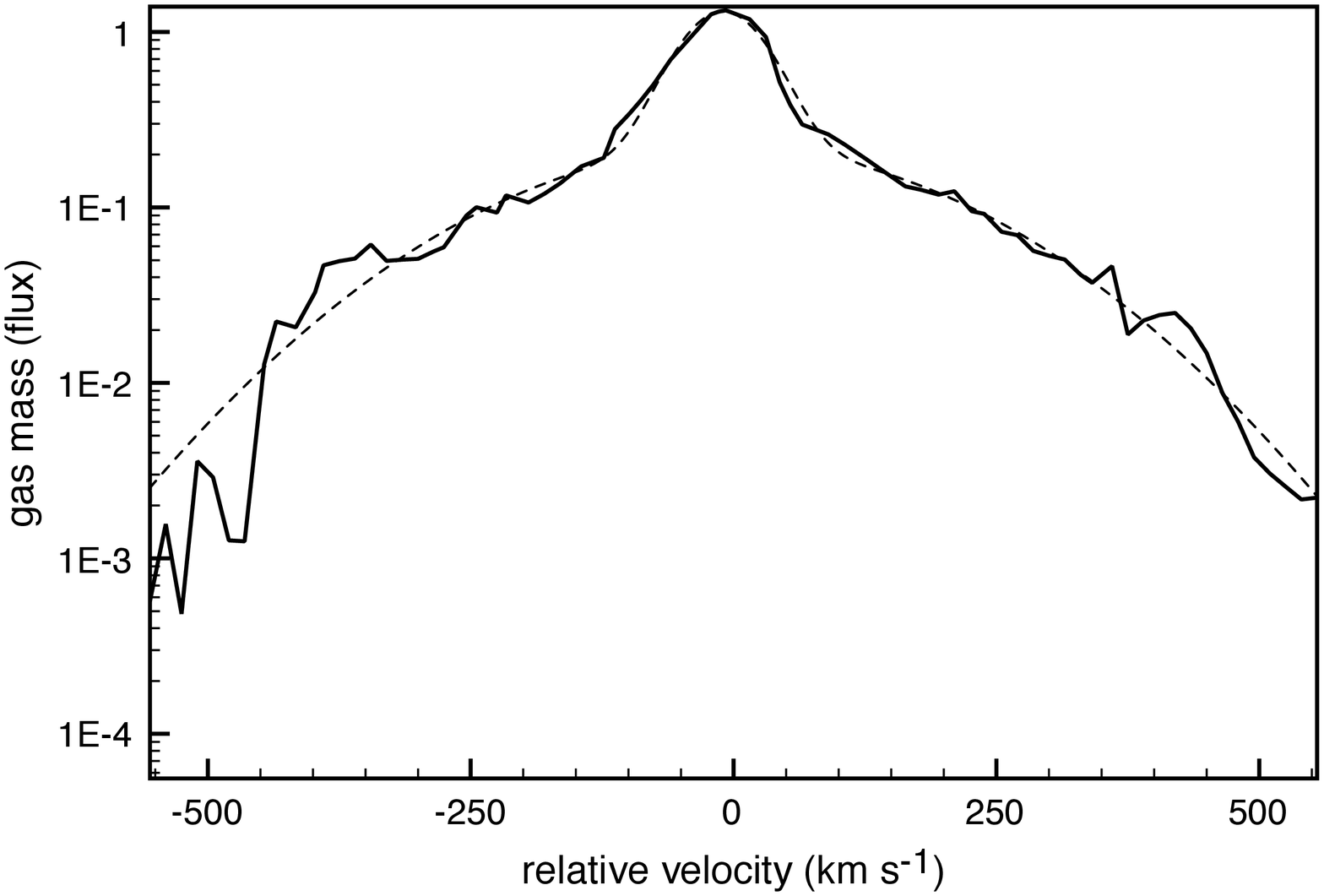}
\caption{\label{spec_stack} Median stacked spectrum comprising all the studied clumps (top: linear scale -- bottom: log-scale), after re-scaling each individual spectrum to the same clump mass (see text). The stacked spectrum is well fitted by a double Gaussian model (dashed), according to which the broad component contains 32\% of the gas mass (a significant part of which, but not all, is above the clump escape velocity), 68\% is in the narrow component (bound to the clump). In our simulations, the gas in the broad component is hot ($\geq \, 10^{4-5}$\,K), outflowing gas.
}
\end{figure}

\begin{figure}
\centering
\includegraphics[width=0.47\textwidth]{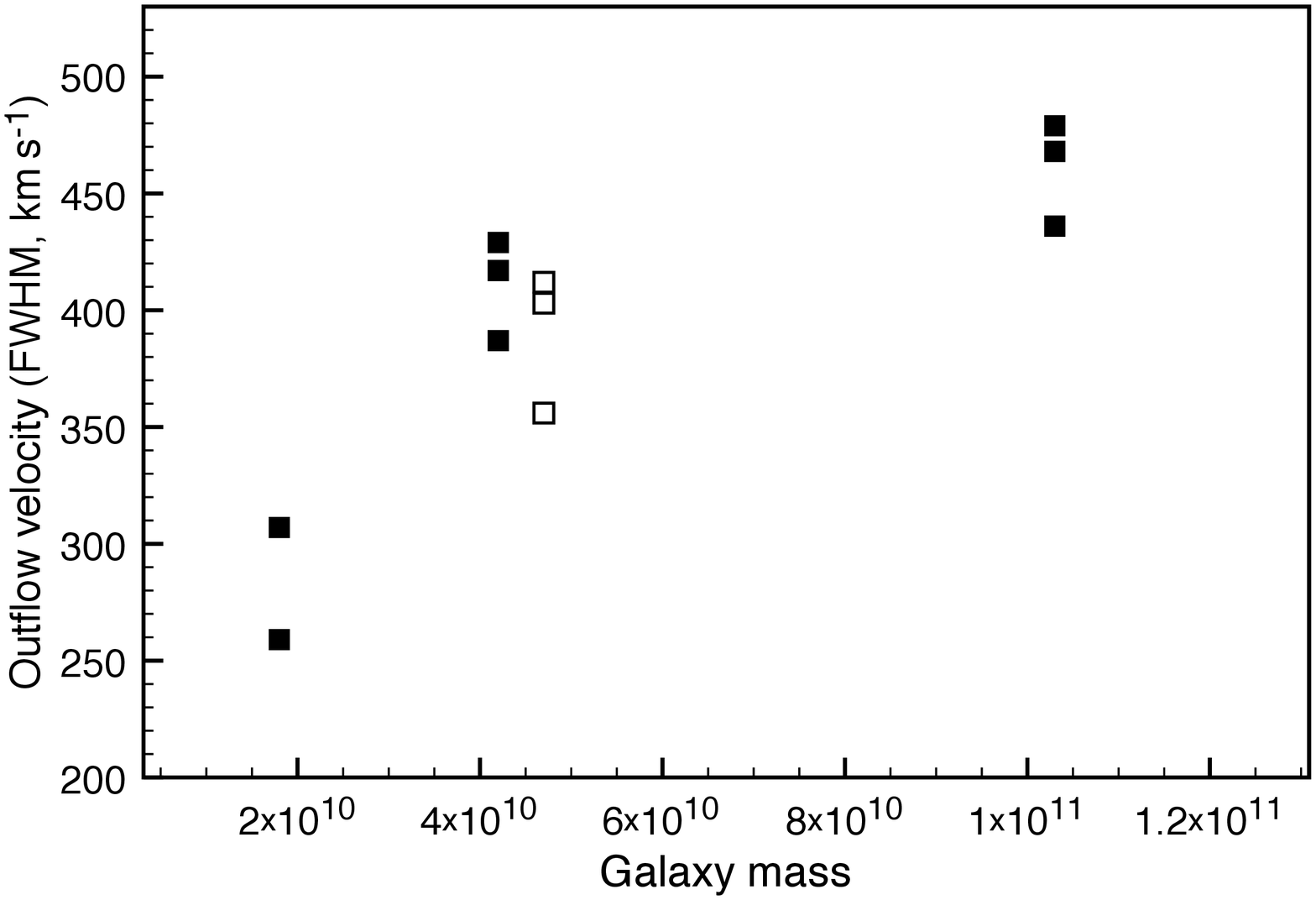}
\includegraphics[width=0.47\textwidth]{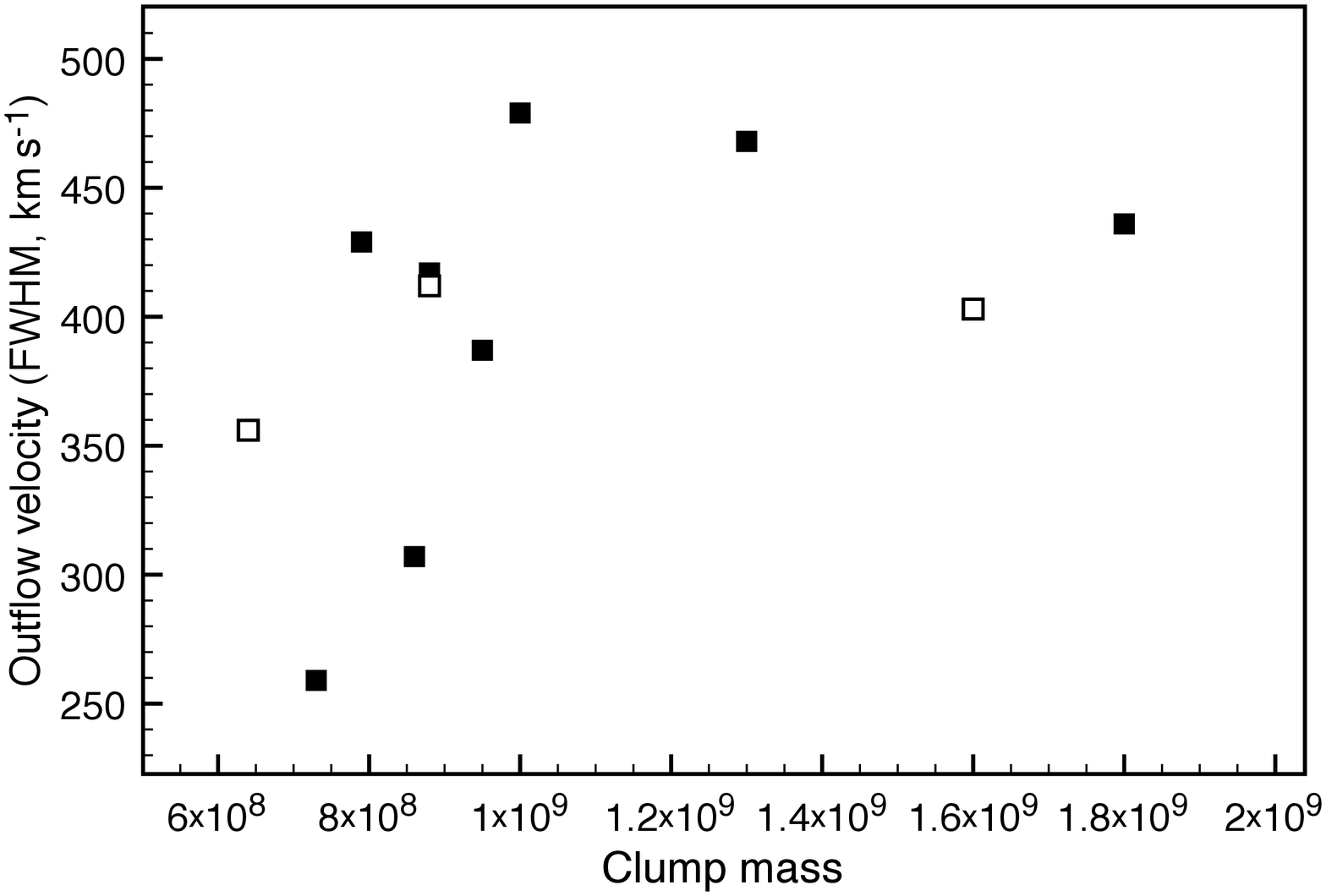}
\caption{\label{scal} Outflow velocity for each clump, measured as the average FWHM of the broad spectral component for each individual clump (spectra are extracted every 40\,Myr and stacked), as a function of galaxy mass (top) and average clump mass (bottom). The open symbols are for model G'2. }
\end{figure}

We show in Figure~\ref{spec_2C} the mass-weighted point distribution function of the gas velocity along a line-of-sight crossing the clump center, using a circular aperture of 300~pc radius around the line-of-sight, and with a line-of-sight inclined by 20 degrees from the disk axis. This is equivalent to a pseudo-observed ``spectrum'' of the clump with a quasi-face-on orientation of the galaxy and a 600~pc FWHM beam\footnote{equivalent to 0.07~arcsec FWHM at $z\,\approx\,2$.}, using crude linear conversion of gas mass into flux. For a typical, quiescent phase, the spectrum shows a narrow component (FWHM$\approx$80\,km\,s$^{-1}$) and a broader component, larger than the clump circular velocity. When examined after a peak in the local star formation rate (due to the swallowing of surrounding gas by the clump), a broad and irregular spectrum is obtained, corresponding to the more disturbed morphology of the same clump on the fourth panel of Figure~\ref{seq_2C}.

We performed the same exercise for all clumps in our samples and display the median stacked spectrum on Figure~\ref{spec_stack}, after re-scaling the velocities by a factor $\sqrt{10^9/M_c}$ where $M_c$ is the mass of the clump in solar masses, i.e. normalizing the baryonic mass of clumps to $10^9$\,M$_\odot$ at all times. The characteristic stacked spectrum displays a narrow component and a broad component, well-fitted by the sum of two Gaussians of respective widths 93 and 424\,km\,s$^{-1}$ FWHM. This velocity widths are close to the circular velocity\footnote{The FMHW of the narrow component is, in detail, somewhat lower than the circular velocity, which comes from the fact that these spectra were obtained for quasi-face-on galaxy orientations, and the giant clumps tend to be relatively aligned with the galaxy disk \citep{ceverino-2}, which reduces the apparent face-on velocity amplitude.}, and above the local escape velocity, respectively, for a normalized clump mass of $10^9$\,M$_\odot$. The best fit for the stacked spectrum is obtained with 32\% of the gas mass in the broad component.

We can confirm in our simulations that this broad component in the spectrum of clumps corresponds to a gas outflow, rather than infall. First, they are obtained with quasi-face-on orientation so there is little fuel for infall onto clumps along the line-of-sight in our idealized experiments. Second, gas in the broad component has temperatures if the $10^4-5\times 10^6$\,K range, indicative of heating by stellar feedback processes. Third, the broad component {\em above} $\sim\,200$\,km\,s$^{-1}$ typically contains $7\times 10^7$\,M$_\odot$ of gas mass, which is able to escape from a kpc-sized clump in $\sim \, 10^7$\,yr, consistent with our direct measurements of outflow rates at a few solar masses per year. This lends support to the idea that similar broad components in observed spectra of high-redshift giant clumps (e.g. Shapiro et al. 2009, Genzel et al. 2011, Newman et al. 2012a) would be most likely attributable to outflows.  The velocity of the gas outflowing from the clumps scales with both clump mass and galaxy mass (Figure~\ref{scal}), with a tighter relation when considering the entire galaxy mass. This shows that giant clumps can launch gaseous outflows that are rapid enough to escape the galactic potential well, even in high-mass galaxies. As previously said, more massive galaxies have more numerous and more massive clumps, and hence the outflows initiated in the giant clumps can reach higher velocities there. 

It is interesting to note that the observed velocity of gaseous outflows in Newman et al. (2012a), which are known to be largely launched by giant clumps (Genzel et al. 2011), scales with galaxy mass, which appears to be consistent with our results. More generally, the global properties of the outflows in our simulations (wind velocity, outflow rate compared to SFR, density range) are quite consistent with existing observational constraints at $z\, \approx 0.5 - 2$ \citep{kornei,rubin,martin}. These observations target high-redshift star-forming galaxies independent of clumpiness, but as a matter of such these galaxies are generally very gas-rich, violently unstable, with significant clumpiness in 50--70\% of cases in the optical and near-infrared (Elmegreen et al. 2007, Mozena et al. in preparation, Guo et al. in preparation). Our simulations are even consistent with the observed dependance of outflow velocities on galactic mass \citep{bordoloi}. Hence our feedback model produces realistic outflows, although the outflow velocity or mass loading are not imposed {\em a priori} in our feedback modeling, and indicates the star-forming clumps can be efficient launching sites for the outflows without being rapidly disrupted.

\subsection{Dynamical loss of aged stars}
The release of material by a clump is not limited to gas outflows. Tracking individual star particles throughout the simulations, we actually measure that a large fraction of stars captured by a clump during its initial collapse, or formed in-situ in the clumps, can gradually escape from the clump in a few $10^8$~yr. 

Independently from any feedback-driven gas outflows, it was already noted by \cite{E05} that typical clumps are not tightly bound compared to the tidal field of their host galaxy, with densities only a factor ten above the limiting tidal density, so that their outer parts could be significantly affected by tidal stripping. Previous simulations did find that stars formed in a clump can gradually escape the clump because of their increasing velocity dispersions over time and because of the galactic tidal field (see for instance BEE07), and the released stars were proposed to fuel the thick disk \citep{BEM09}. The same effect is present in our simulations, in somewhat larger proportions, presumably because the gas outflows both regulate the growth of the clump mass and increase the stellar escape rate from the shallower local potential well.

To quantify the dynamical loss of stars by clumps, we measure the mass of stars present in the clump at a given time and that have left the clump later-on\footnote{We subtract from this quantity the mass of stars lying outside the clump at the former instant and found in the clump and the latter one, which means that the measurement is corrected for potential rapid entry/re-escape of pre-existing background stars from the large-scale disk, which is found to be a relatively minor quantity.}. The fact that the obtained quantity is always positive confirms that it corresponds to the loss of in-situ material and is not significantly affected by chaotic entry and escape of stars formed elsewhere. This typical ``stellar loss rate'' is 0.2--2\,M$_\odot$\,yr$^{-1}$ for giant clumps, somewhat lower but of the order of their internal star formation rate. The clumps can release aged stars almost at the rate at which they form new stars.

\begin{figure*}
\centering
\includegraphics[width=0.42\textwidth]{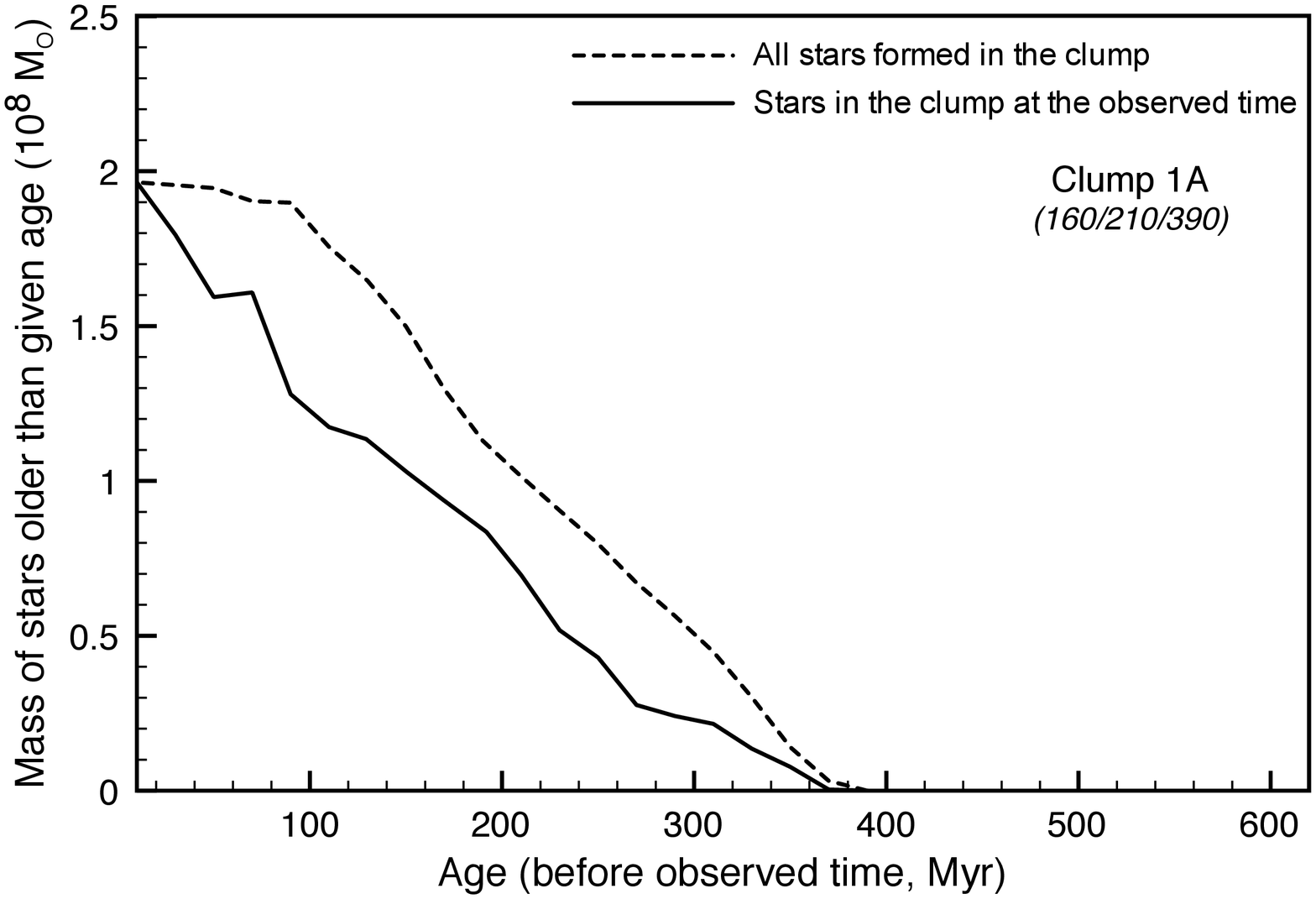}
\includegraphics[width=0.42\textwidth]{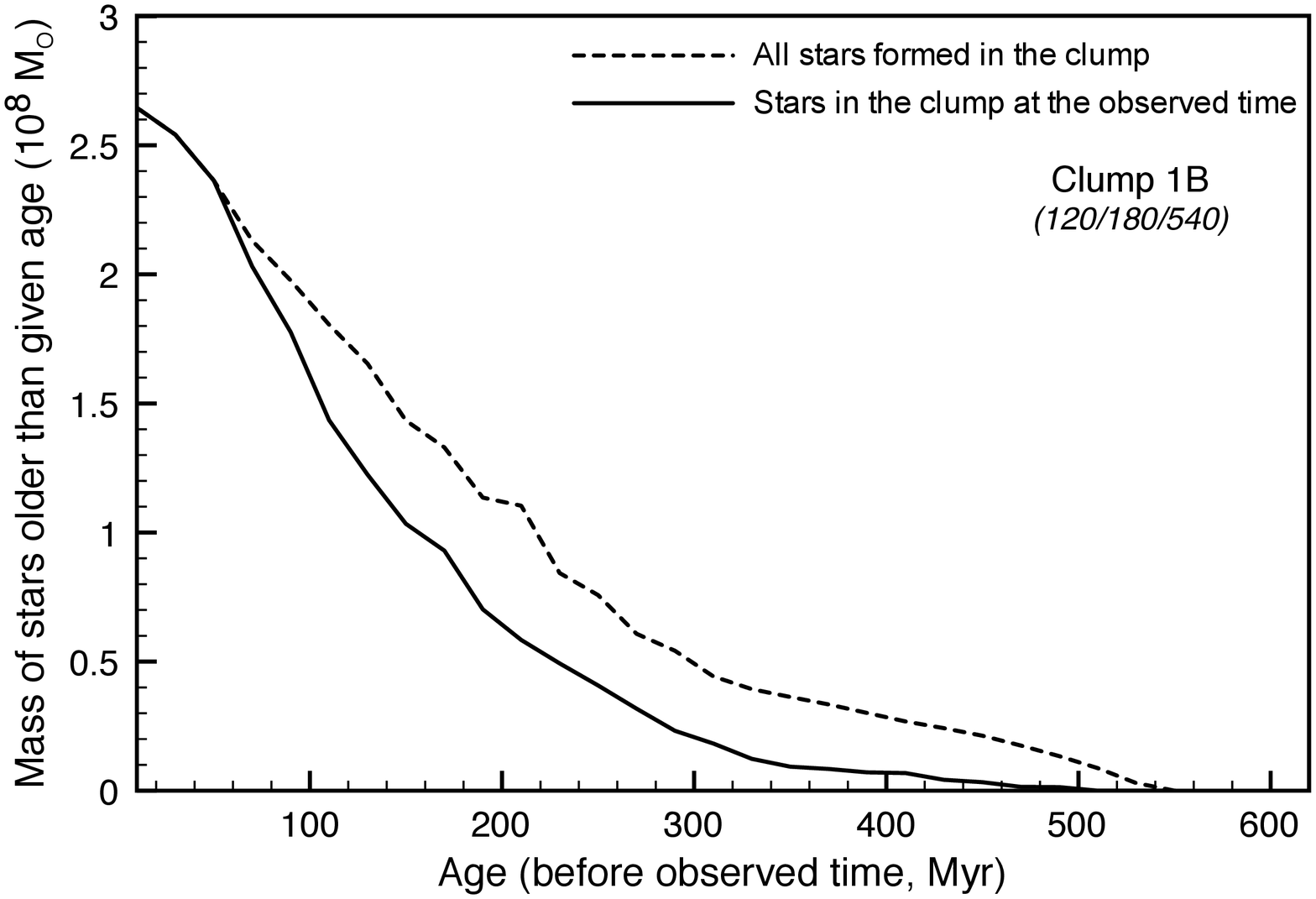}
\includegraphics[width=0.42\textwidth]{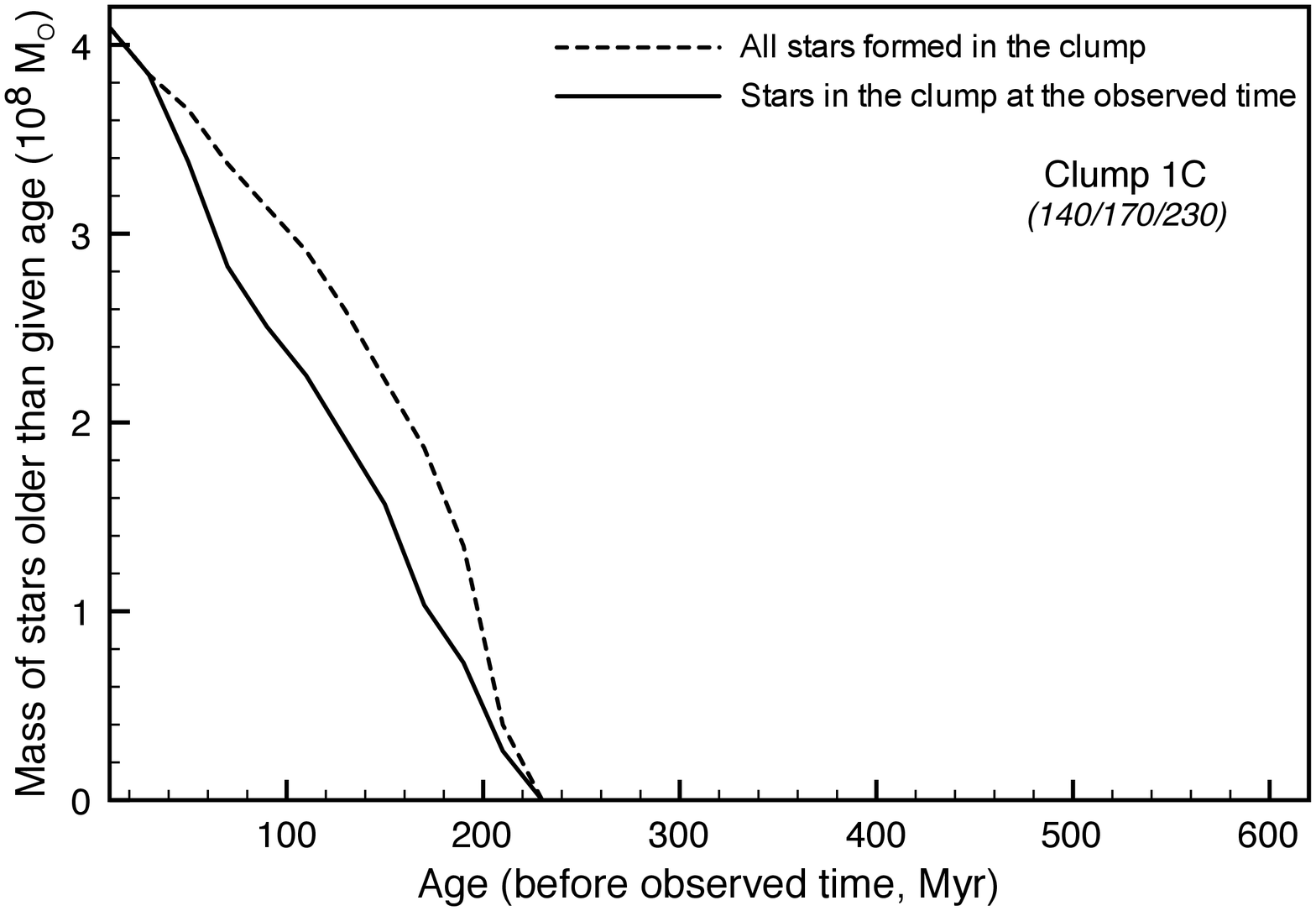}
\includegraphics[width=0.42\textwidth]{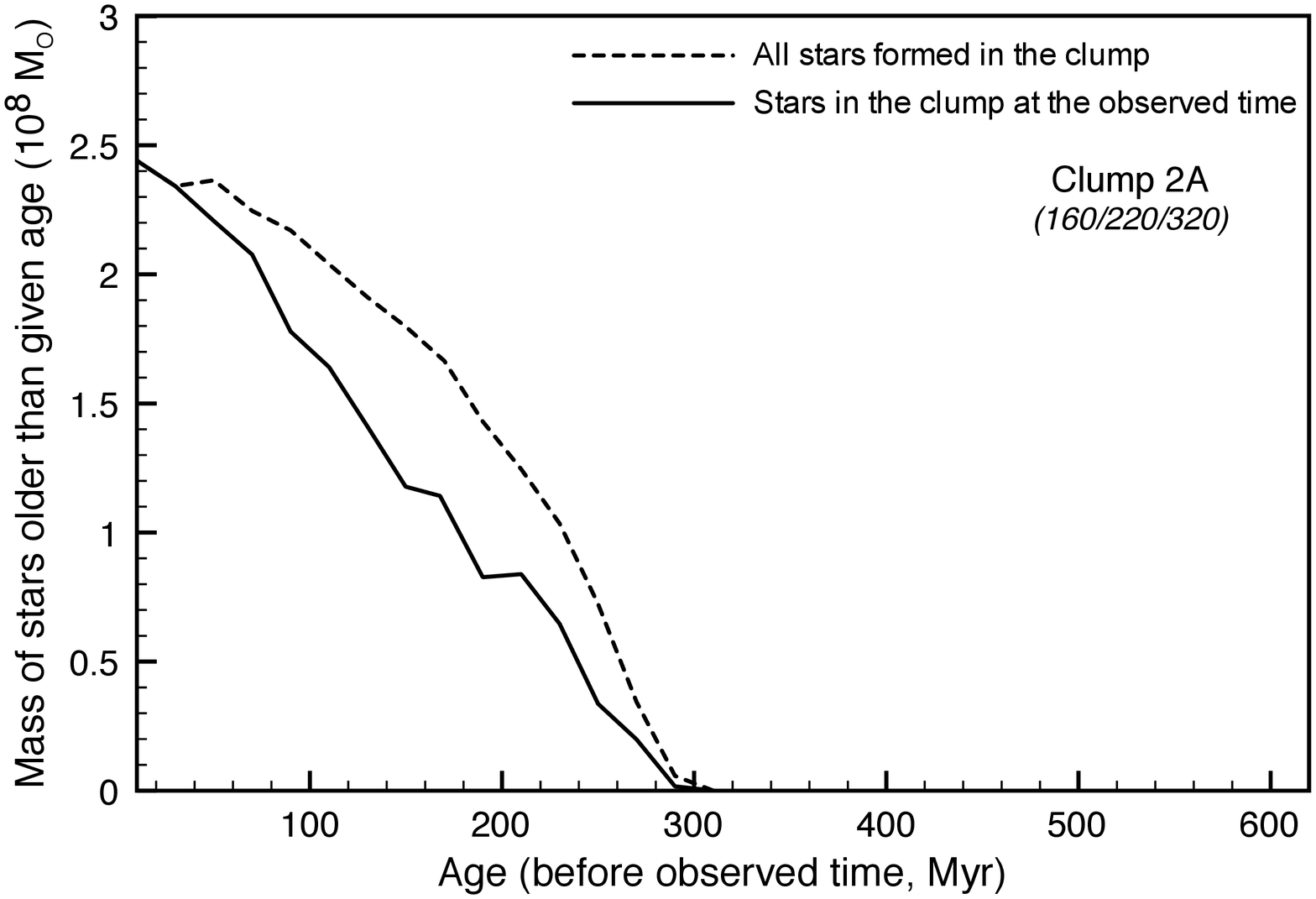}
\includegraphics[width=0.42\textwidth]{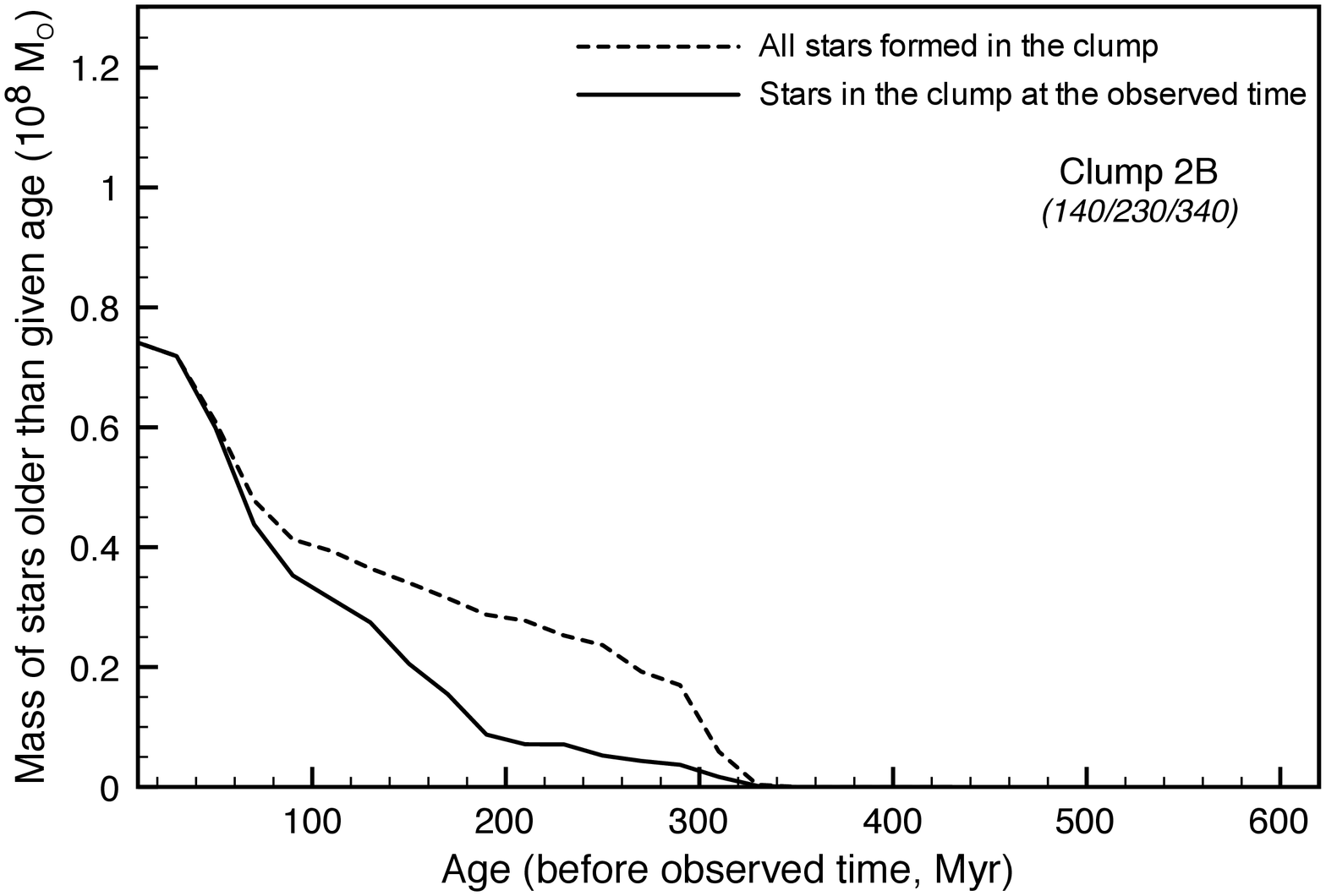}
\includegraphics[width=0.42\textwidth]{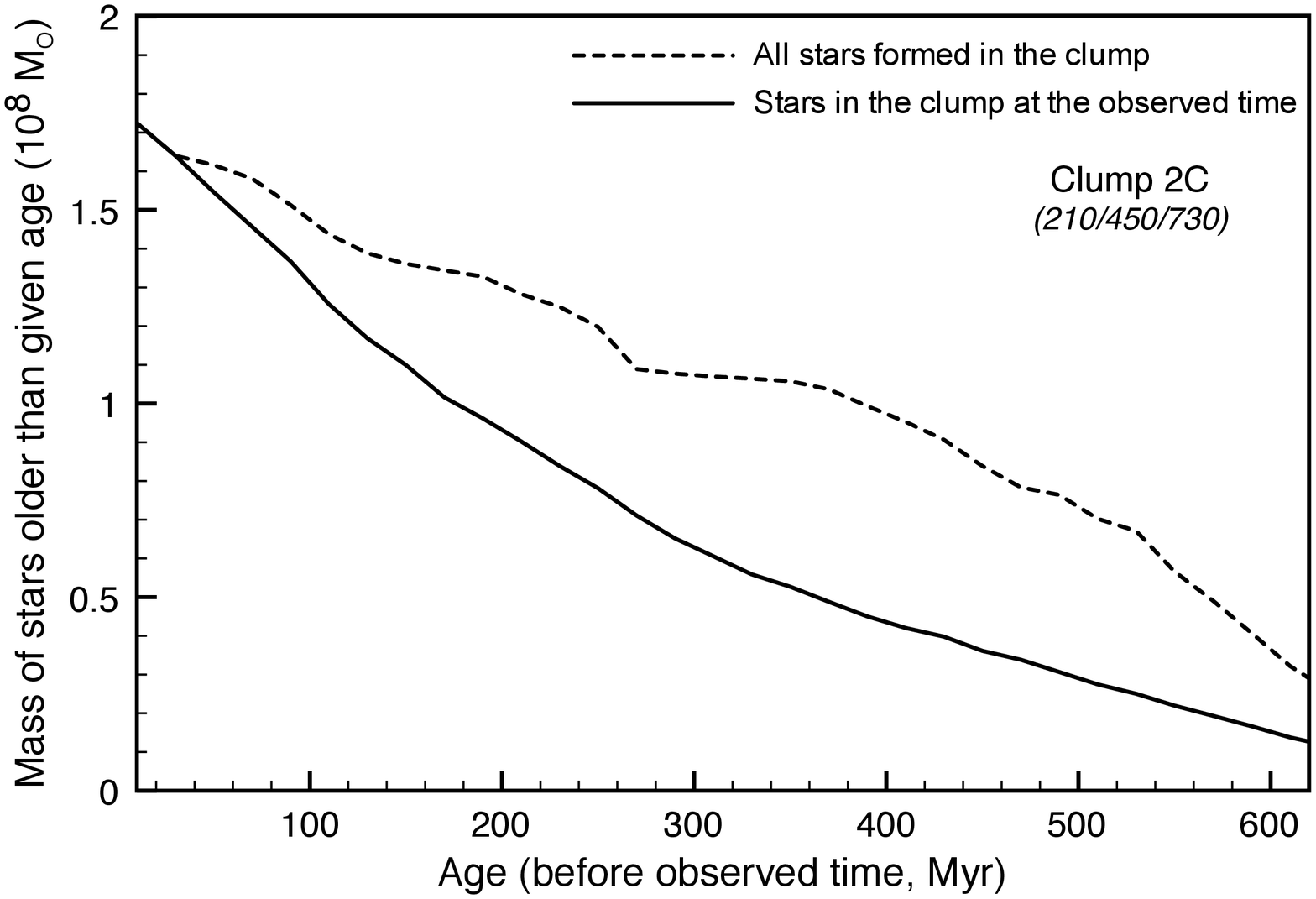}
\includegraphics[width=0.42\textwidth]{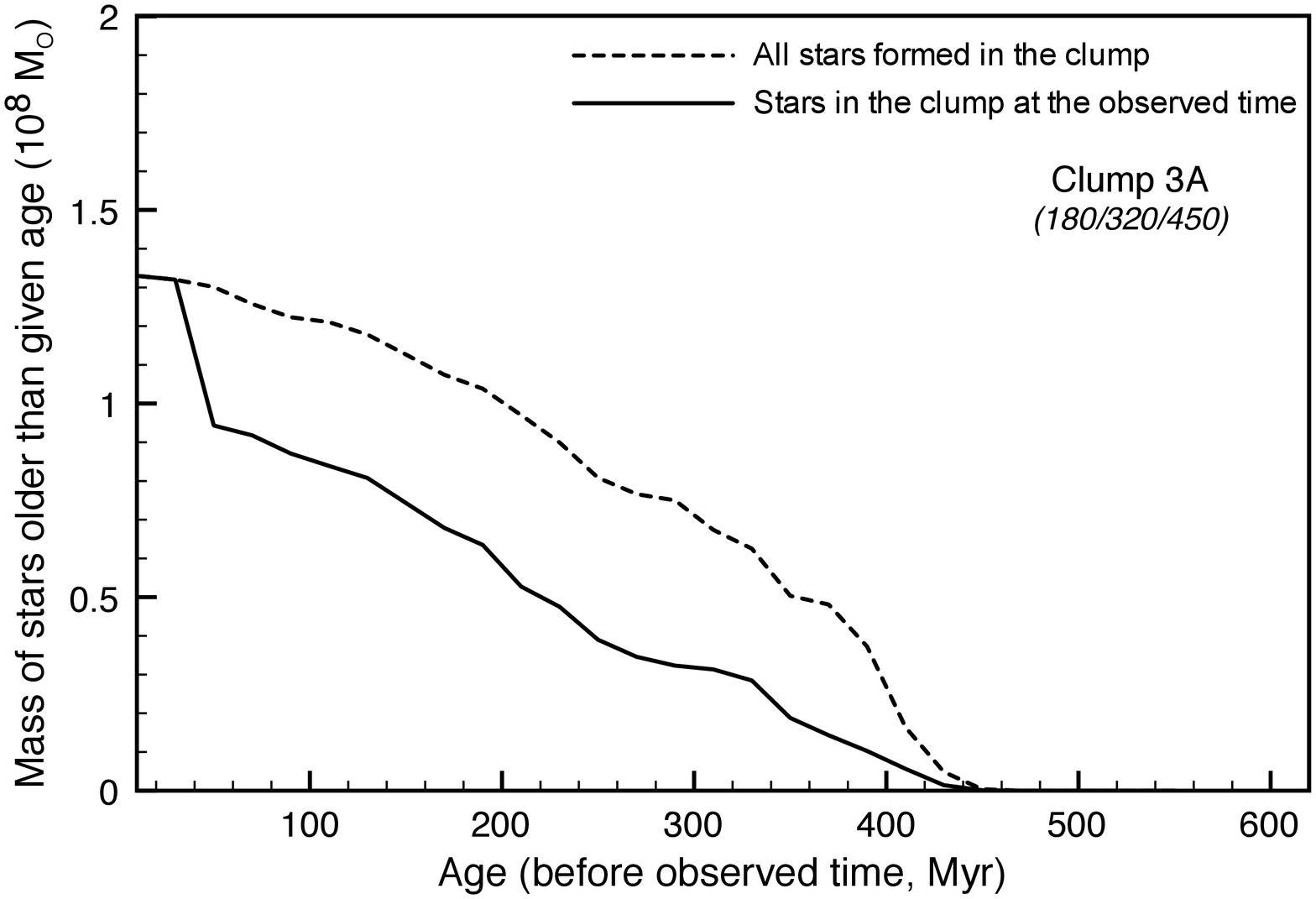}
\includegraphics[width=0.42\textwidth]{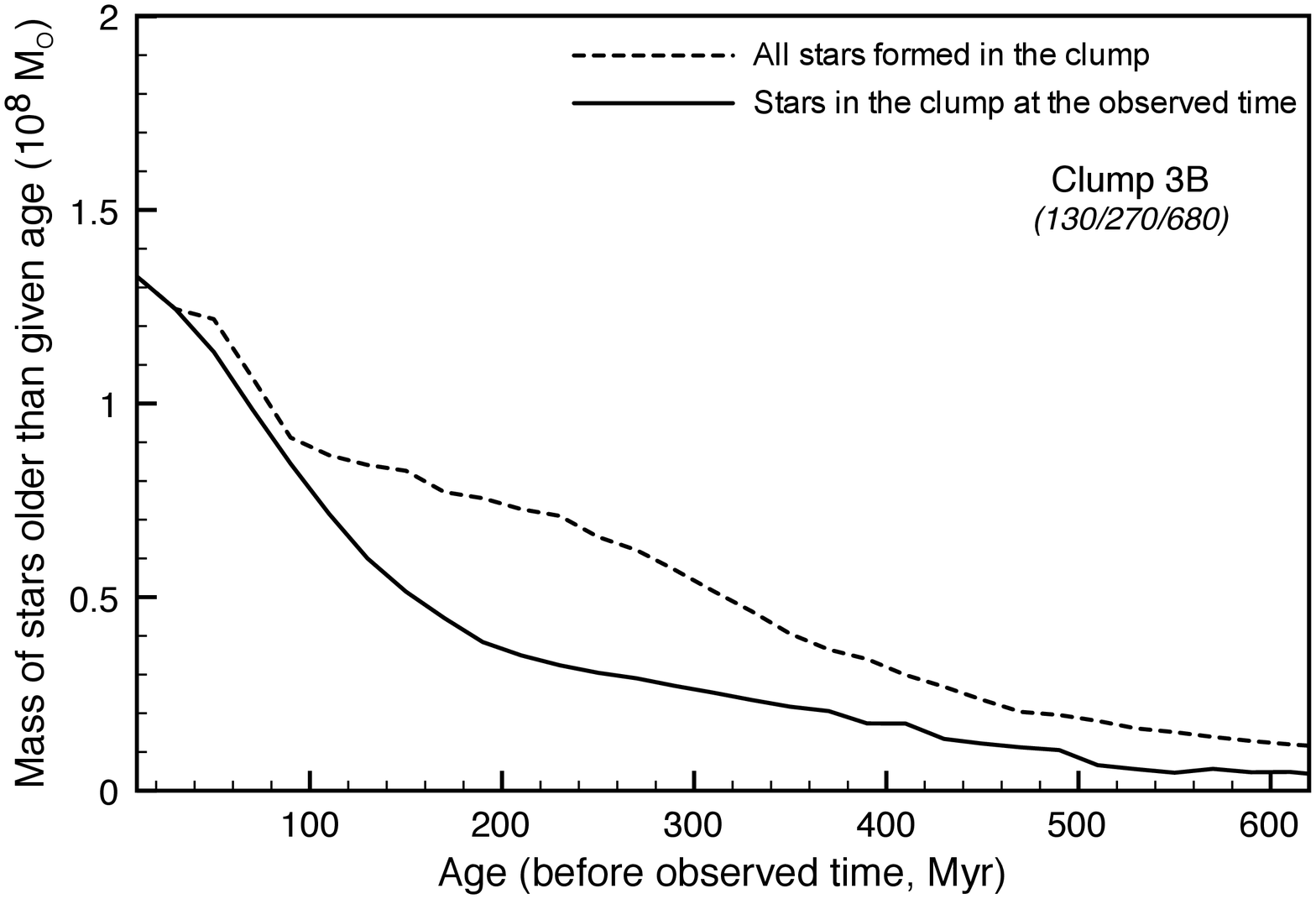}
\caption{\label{pop_ages} Cumulative age distribution of stars present in clumps, for a selection of relatively old clumps (each of the clump tracked in our models is here analyzed when it is about to enter the central kiloparsec). The dashed line shows for comparison the cumulative age distribution of all the stars that formed in each given clump, regardless of their location (in or outside the clump) at the analyzed instant, and normalized to the same final value for clarity. Three numbers are given in Myr for each case: median age of the stars that lie in the clump at this instant / median age of the stars that have formed in the clump (but may not lie in the clump anymore) / actual age of the clump main progenitor as tracked in the simulation. The typical age of stars in an evolved clump is rarely larger than 200\,Myr, even for clump ages of 300-600\,Myr and more.
}
\end{figure*}

\begin{figure}
\centering
\includegraphics[width=0.47\textwidth]{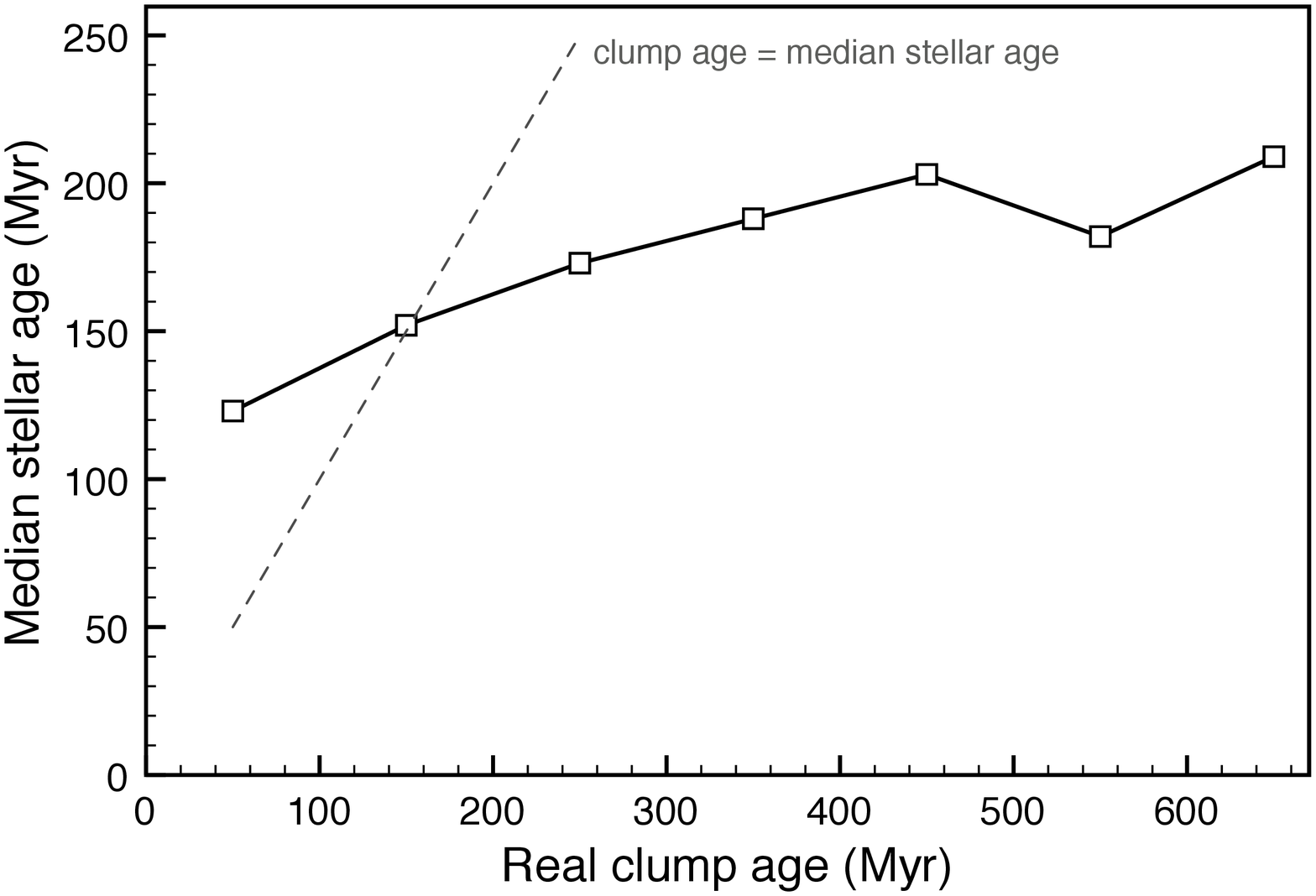}
\caption{\label{ages} Median age of the stars contained in the clump, as a function of the clump age. We show the average measurement, for all clumps in each age bin of 100\,Myr. Clumps of 500\,Myr and more contain stars of typical age $\leq$\,200\,Myr, because of the dynamical loss of older stellar populations. Young clumps of 100\,Myr and less can have an average stellar age larger than the actual clump age, because they capture pre-existing stars during their initial collapse. We assumed random ages from 0 to 1\,Gyr for the stars present at the beginning of the simulation.}
\end{figure}

\bigskip

To quantify the effect on the observable stellar ages of the clumps, we examine relatively old clumps in our simulations. For each of the eight clumps tracked in our fiducial runs, we ``observe'' it at the last snapshot before it enters the central kiloparsec, i.e. at a relatively advanced stage -- the idea being that young clumps cannot have their stellar content severely affected by the long-term loss of stars. The ``aged clumps'' picked this way have ages ranging from 230 to 730\,Myr, on average 400\,Myr. Then, for each of these clumps, we identify the stars that have formed in the clump at any time, independently of their final location. We find that {\em the median age of stars that have formed in a given clump is about half of the real age of the clump} (Figure~\ref{pop_ages}, dashed curves). This is because the clumps maintain a relatively constant star formation history owing to external gas accretion. Then, we identify the stars that are present in each clump at the ``observed'' time, and we find that because a higher fraction of aged stars have escaped from the clump, {\em the median age of stars lying in a clump is less than half of the real age of the clump} (Figure~\ref{pop_ages}, solid curves). Our sample of ``aged clumps'' have a median real age of 420\,Myr (ranging from 230 to 730\,Myr), but the age of the stars that they contain has a median of 150\,Myr (ranging from 120 to 210\,Myr).

Of course the effect is weaker for younger clumps. We show in Figure~\ref{ages} the median age of the stellar populations contained in a clump as a function of the real clump age, as gathered from our entire sample. The median age of stellar populations saturates around 200\,Myr even for older clumps, because the non-declining star formation rate is comparable to the gradual dynamical loss of older stars. Note that clumps do capture pre-existing stars during their initial collapse, as expected for a two-fluid instability \citep{Fellhauer06,Elmegreen11}, and thus very young clumps can have a median stellar age older than the clump itself, as also seen in Figure~\ref{ages}.

\begin{figure*}
\centering
\includegraphics[width=0.86\textwidth]{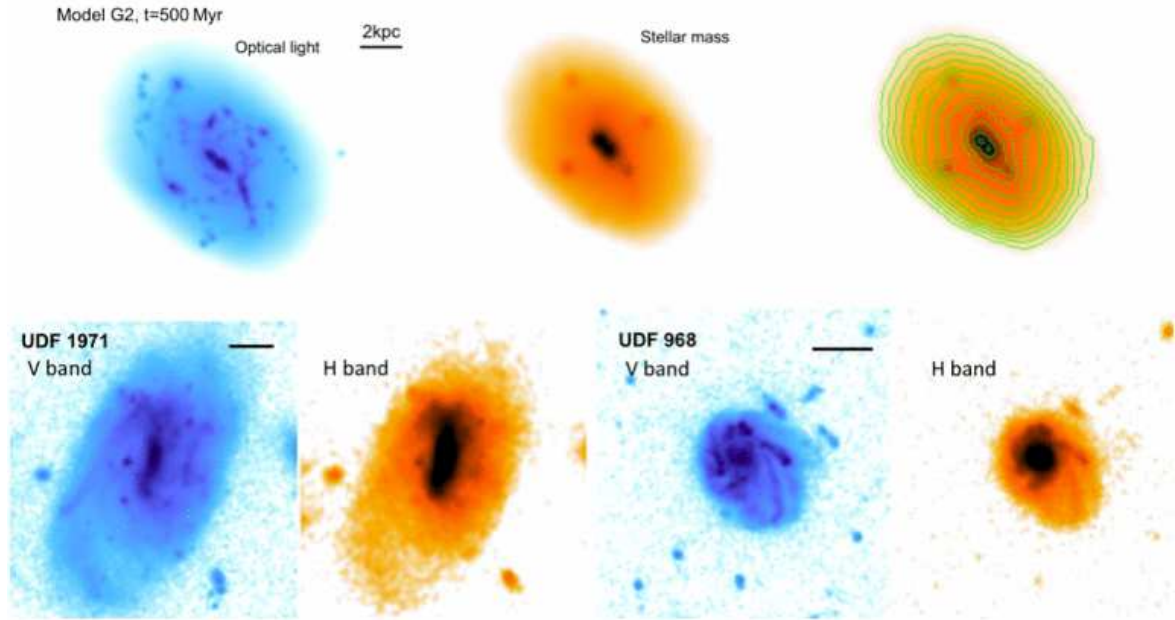}
\caption{\label{star_mass_map} 
Stellar mass map and optical light image of galaxy G2 at $t\,=\,500$\,Myr. The optical light image is generated assuming that the mass-to-luminosity ratio of stellar populations is constant during 10\,Myr, and subsequently decreases as $t^{-0.7}$. The contrast clump in stellar mass maps is much lower than in optical images or gas density maps. For instance, the clump visible to the top is clump 2C, which has formed 400\,Myr ago in the outer disk. On the right panel, logarithmic isodensity contours with a spacing factor of 1.5 between contours are overlaid (i.e., the level of a given contour is 1.5 times the level of the previous one). It shows that on such stellar mass maps, the peak surface density of a clump is only 1.5-2 times that of the surrounding material. Here we applied Gaussian smoothing with FWHM of 200\,pc, lower than the clump size, and no noise was added, so the low contrast of the clumps is not an effect of dilution at low resolution. Note that the typical resolution of HST/WFC3 imaging may lead to some extra decrease in the clump contrast because of lower-resolution beam smearing. On the bottom, we show the HST/ACS and HST/WFC3 observations of two clumpy galaxies from the Elmegreen et al. (2007) sample, UDF\,1971 and UDF\,968. The clumps have a low contrast in the near-infrared but the optical images are much more clumpy with giant clumps, smaller clouds, some spiral features and offset centers, similar to the advanced stages in our clumpy galaxy simulations. Note that the small knots seen around these two UDF galaxies are not minor mergers or satellites, but background galaxies (at least for most of them, the photometric redshift is much larger).
}
\end{figure*}

We also learn that the stellar mass of clumps is regulated, as the rate of new star formation is not much larger than the stellar escape rate, and thus the stellar mass in a given clump does not strongly increase over time. The average stellar mass fraction in the clumps at 100-200\,Myr of age is 18\% (the remaining 82\% is gas), and it only marginally increases to 23\% in the 400-600\,Myr age bin. Hence the contribution of clumps to the stellar mass distribution is weaker than their contribution in the spatial distributions of gas and star formation. As illustrated on Figure~\ref{star_mass_map} the contrast of clumps w.r.t surrounding disk material in stellar mass density maps is low, typically less than a factor two. The clumps are much more contrasted in the equivalent optical light image (in which recently-formed stars are more prominent), which compare favorably to real galaxies at high redsfhit (Figure~\ref{star_mass_map}).

\subsection{Accretion rate onto the clumps}
Comparing the gas mass of clumps and the measured outflow rate, we infer the rate of gas accretion by the clumps, which would be hard to measure directly given the non-Lagrangian nature of the gas component in our simulations. The result is shown for several clumps in Figure~\ref{rate_curves}. The obtained value is sometimes negative, which traces the uncertainty in other quantities and/or the potential tendency to slightly under-estimate the outflows\footnote{As explained in Section~3.2, we chose to measure the outflowing component perpendicular to the disk, the outflow around spherical boundaries being somewhat larger, but consistent. This choice was made to be conservative in the estimated outflow rates. The negative values of the estimated gas accretion rate onto the clumps can thus correspond to gas that leaves the clumps within the disk plane, either because of in-plane feedback-driven outflows, or because of tidal stripping of the gaseous component.}. The average of the positive values is five times larger than the absolute average of the negative values, confirming that the uncertainty is relatively small, and that we are indeed measuring net inflow.

The gas accretion rate by the clumps is typically 2--15\,M$_\odot$\,yr$^{-1}$, with the peak values corresponding to the accretion of smaller gas clouds (some of which are visible in the simulations snapshots) while the lower background values correspond to continuous accretion of diffuse disk material. This accretion rate is higher during the first 50\,Myr, when each clump is initially collapsing (Fig.~\ref{rate_curves}), then it fluctuates  about a level of a few M$_\odot$\,yr$^{-1}$.

The accretion rates measured in our simulations are consistent with those predicted analytically by \citet{DK13}. The predicted rate of capture of gas by a massive clump in a disk is $\alpha \, \rho_d \, R_{\rm tid}^2 \, \sigma_d$ with $\alpha \approx 1/3$ \citep{DK13}, where $\rho_d$ is the typical density of gas in the disk, $\sigma_d$ the gas velocity dispersion, and $R_{\rm tid}$ is the tidal radius beyond which gravitational forces from the galaxy dominate w.r.t the internal gravity of the clump. We apply this to galaxy G2, where the measured gas velocity dispersion is $\sigma_d=44$\,km\,s$^{-1}$, and the gas density in the mid-plane at the half-mass radius is 2.3\,M$_\odot$~pc$^{-3}$. The average mass of clumps 2A, 2B and 2C (averaged over the clumps and over time) is $7.8\times 10^8$\,M$_\odot$. The tidal radius at which forces from the galaxy (of mass\footnote{including the mass of dark matter inside the galaxy half-mass radius} $1.2\times10^{11}$\,M$_\odot$) and from the clump are roughyl equal\footnote{if we approximate both the clump and the galaxy with two point-masses on a circular orbit} is $R_{\rm tid} \approx \sqrt{7.8\times 10^8 / 1.2\times10^{11}} \times R_{\rm gal}$, where $R_{\rm gal}$\,=\,3.3\,kpc is the half-mass radius of this galaxy. This gives $R_{\rm tid} \, \simeq \, 266$\,pc, so the predicted average accretion rate onto the clumps is about 7\,M$_\odot$\,yr$^{-1}$. The average measured value in our simulation, after the first 100\,Myr during which the clumps are collapsing with high infall rates, is 5.3\,M$_\odot$\,yr$^{-1}$, very consistent with the simplified analytic estimate provided by Dekel \& Krumholz (2013).

% - - - - - - - - - - - - - - - - - - - - - - - - - - - - - - - - - - -

\section{Discussion}

\subsection{Observed outflows and stellar populations}
The recently detected outflows in giant clumps at $z \, \approx \,2$ have gas outflow rates of $\approx$\,1-2 \,$\times$\,SFR, with perhaps extreme cases up to $8\,\times$\,SFR (Genzel et al. 2011, Newman et al. 2012a). About 60\% of the clumps in our models are in this range. Given that accurately resolving the clumps requires high sensitivity and adaptive optics, the observed samples remain very small and mostly consist of the brightest clumps in the most massive disk galaxies: their individual mass and SFR are typically higher than in the more ``typical'' clump selected in our models, which might lead to different trends in the outflow rate to SFR ratio (stronger activity, but deeper potential wells). Given these limitations, and in spite of the strong agreement between our fiducial models and observations, it is probably premature to assume that our alternative model with reduced SN feedback, which typically produces outflow rates somewhat lower than the SFR, is less realistic. Nevertheless, the comparison of theis model with our fiducial ones shows that supernovae do play a role in clump-scale outflows, the rate of which is not determined solely by the momentum injection through other feedback modes. 

Our simulations are also in agreement with the observed stellar populations. The continuous accretion of mass by the clumps keeps their gas mass about constant, and their star formation histories are relatively flat (with large fluctuations, but without a significant systematic trend). Wuyts et al. (2012) indicate a typical age for stellar populations of 100--200\,Myr, assuming exponentially-decaying star formation histories. This assumption can somewhat under-estimate the age of stellar populations with more constant SFR histories \citep{maraston}. The continuous re-formation of young stellar populations in the clumps, coupled with the gradual tidal loss of aged stars, make the clumps of real age 300\,Myr and above appear as containing a median stellar age of about 200\,Myr. This effect, already noted\footnote{The effect was somewhat weaker in the Bournaud et al. 2007 models, presumably because the lack of gaseous outflows maintained a deeper potential well that could retain the stars on longer timescales} in Bournaud et al. (2007), puts our simulations in agreement with the typical observed stellar ages in the clumps.

\subsection{Short-lived clump models}
Other numerical simulations of high-redshift clumpy disks have used strong momentum-driven feedback to model short-lived clumps (Genel et al. 2012, Hopkins et al. 2012). In addition to using a high trapping factor for photons, providing high momentum to the gas, these models assumed that the initial outflow velocity should be equal to the local escape velocity from the galaxy or from the clump (multiplied by a scaling factor close to unity). This hypothesis maximizes the deduced mass loading of the outflows : for a given momentum $m \times v$ this puts the highest possible gas mass $m$ above the escape velocity. The observed wind velocities may actually me larger (Genzel et al. 2011), which would reduce the involved mass for a given amount of momentum. In addition the hydrodynamic interaction was also suppressed over some distance in the Genel et al. (2012) models, preventing dilution of the outflow momentum and thus boosting up the mass loading. In contrast, in our simulations we evaluate the outflow properties of velocity and mass loading from the physical elements of the feedback processes, and we find that some components in the outflows are much more rapid than the escape velocity (up to 500\,km\,s$^{-1}$ in agreement with data), in which case the resulting mass loading is lower for the available momentum. Conversely some regions in the outflows have moderate velocities (around 100\,km\,s$^{-1}$) and remain eventually bound in the clump's potential well unless they exchange momentum with other components.

These other simulations with very extremely strong feedback have demonstrated that it is possible to find combination of star formation efficiencies and feedback schemes that allow clumps to gather about $10^9$ solar masses of baryons for a short time before clump disruption. However, the arbitrarily strong feedback imposed disrupts the clump within typically 30-50\,Myr, which appears very short compared to observed stellar populations ages up to 200\,Myr, and other clumps do not re-form on an equally-short timescale. Hence, it is rather unlikely that short-lived clump models could account for the high frequency of giant clumps in high-redshift disks and the 200\,Myr-old stellar populations found in these clumps.

\subsection{Long-lived clump models and the necessary regulation of bulge growth}

The inclusion of realistic outflows induces several interesting differences compared to previous long-lived clump models. The baryonic mass of clumps is, on average, constant over their inward migration in the disk. Models without feedback or limited to (weak) supernova feedback found clumps whose mass was steadily increasing along their inward migration, because of the same process of continuous mass accretion as in the present models (e.g., Noguchi  1999, Bournaud et al. 2007, see also the clump growth in the ``no wind'' models by Genel et al. 2012). The continuous mass accretion by clumps can be largely compensated by gaseous outflows\footnote{Stellar loss was present in previous models, but probably weakened by the deeper potential wells associated to the absence of gaseous outflows} in our new models, and as a result the mass of clump when they reach the galactic center is regulated to 0.2--2.4\% (on average 0.8\%) of the galaxy baryonic mass in our current simulation sample. Furthermore the loss of aged stars and re-accretion of gas keeps much of this mass gaseous, with an average gas fraction in the clumps in our sample of about two thirds, and an average stellar mass of $3\times 10^8$\,M$_{\odot}$. 

If a generation of clumps contains 5-10 clumps, each conveying $3\times 10^8$\,M$_{\odot}$ of stars to the bulge in $\approx$\,500Myr, and three such generation spans a galaxy's life in the $z\,\approx3-1$ range, this mechanism overall provides about $7\times 10^9$\,M$_{\odot}$ of stars to a bulge component -- which would be a ``classical'' bulge for a clump coalescence mechanism \citep{EBE08}. This is a rather reasonable mass for galaxies with stellar masses reaching several $10^{10}$\,M$_{\odot}$ by redshift zero. The clumps will also carry gas inward (through their migration if they are long-lived, and through the general instability-driven inflow in any case) but if the gas preserves a significant angular momentum it may form the inner regions of exponential disks, or a disky (pseudo-)bulge \citep[see also][]{inoue1}. Hence there is not necessarily over-production of bulges in such a long-lived clump model. Alternatively, the instability-driven inflow may grow a spheroid in which central star formation is continuously fueled with the appearance of ``blue nuggets'' (Zolotov et al. in preparation, Barro et al. 2013). A detailed budget of bulge growth and disk survival in our current clumpy disk models, and in similar models of high-redshift mergers, will be presented elsewhere (Perret et al. 2013b). 

A large fraction of observed high-redshift clumpy galaxies are already more massive than today's Milky Way at $z\, \approx \,$2, and are not expected to remain disk-dominated by redshift zero. Smaller galaxies, with the typical mass expected for a Milky Way-like progenitor at $z\,\approx\,2$,  also seem clumpy in their morphology \citep[even perhaps down to lower redshifts][]{E07, bournaud12}, and their apparently ``dispersion-dominated'' kinematics are also typical for clumpy disks once studied at high resolution \citep{Newman12b}. It appears in our simulations that such lower-mass galaxies tend to have somewhat lower-mass clumps. Over a 1\,Gyr-long period, our low-mass model G3 forms only four clumps more massive\footnote{staying above $4\times 10^{8}$\,M$_{\odot}$ for at least 40\,Myr} than $4\times 10^{8}$\,M$_{\odot}$, while twelve such clumps are found in the high-mass model G1. Given that lower-mass clouds are easily disrupted by feedback (and that the associated outflows may more easily escape the galactic potential well), the relative contribution to bulge growth should be weaker in low-mass galaxies. This could naturally explain the correlation between stellar mass and bulge fraction along the Hubble Sequence, but should be studied with physically-motivated feedback models in full cosmological context so as to encompass the full duration of clumpy phases.

The bulges that grow in our models, in addition to being reasonably massive, are very gassy, because the clumps evolving with realistic feedback remain gas-dominated -- simulations with weak supernovae-only feedback produce clumps whose mass and stellar fraction were increasing over time. Although we defer the analysis of the bulge properties to a subsequent paper,  we here speculate that these bulges could often resemble pseudo-bulges, since they form as gas-rich central mass concentrations. Relatively stochastic growth of a pseudo-bulge could thus likely take place in these systems. The instability-driven inflow can grow the central Supermassive Black Hole (SMBH) in a more continuous way, as predicted in simulations and supported by the high AGN fraction in clump disks \citep{bournaud12} and the high AGN fraction in high-redshift star-forming galaxies in general \citep{mullaney, juneau}. Steady SMBH feeding along with more stochastic (pseudo-)bulge growth could thus be a path to explain the poor correlation between SMBH mass and pseudo-bulges \citep{kormendy}.

\subsection{Self-regulation of the clump mass}

We have noted in the previous parts that when the star formation rate of a given giant clump increases, its outflow rate also increases, and generally in larger proportions (at least when averaged over time periods of 40\,Myr, Fig.~7). We here suggest that this response can compensate for episodes in which the mass of a giant clumps tend to grow, for instance when a smaller clump is swallowed. The strong subsequent ouflows can prevent growth of the clump mass: without this, the ability of clumps to accrete surrounding material would increase and their mass would continue to increase. 

For a star formation rate SFR\,=\,$S$ expressed in M$_\odot$\,yr$^{-1}$, the outflow rate can be written as $S^q$ and the best fitting exponent in our models is found\footnote{best fit for the star formation rate and gas outflow rate averaged over 40\,Myr time bins} to be $q \approx 2.2$. This is a crude description of the feedback behavior, but it captures the fact that moderate increases in the SFR can turn into larger increases in the outflow rate, as seen in the individual tracks on Figure~\ref{rate_curves}. In a steady state the accretion rate should be $A \approx (1+\epsilon_{\rm escape}) S + S^q $ to keep the clump mass constant, where $\epsilon_{\rm escape} \approx 0.5$ if the fraction of stars formed in the clump that eventually escape the clump (after a time delay of the order of 200\,Myr in our simulations, neglected here). A clump forming in a non-turbulent medium would have a mass accretion rate equal to the Jeans mass divided by the gravitational free-fall rate, during its initial collapse phase. This quantity is $A = \sigma^3/G$. Since the giant clumps of high-redshift galaxies move rapidly in a turbulent disk, the surrounding reservoir is continuously replenished at the ambient disk density, keeping this accretion rate constant. The steady state condition is thus $ (1+\epsilon_{\rm escape}) S + S^q  = \sigma^3/G$ with $q=2.2$. Hence a typical clump forming stars at a rate of 2\,M$_\odot$\,yr$^{-1}$ can live in steady state if its binding velocity dispersion is 38\,km\,s$^{-1}$, which is indeed close to the measured dispersions in our models (Section~3.1), and in the molecular gas of high-redshift galaxies. The somewhat higher dispersions measured in the simulations (38--53\,km\,s$^{-1}$) may result from the tidal stirring of clumps, otherwise the excessive binding would lead to a steadily increasing clumps mass. Modeling the feedback response with a power law as done here is of course arbitrary, but the global idea is that there is a steep response where a moderate increase in the accretion rate and SFR can be followed by a stronger increase in the outflow rate (ass previously see in Figure~\ref{rate_curves}), leading to self-regulation of the clump mass.

%\subsection{Globular cluster formation}
%{\em to be completed}
%The constant, fluctuating SF histories, regenerated by external accretion on clumps after the initial collapse+burst of the clumps, is exactly what is required to explain the stellar populations in GCs. So if substructures in the giant clumps can become GCs (we do see such substructures in higher-resolution simulations with shorter durations) they can have the required stellar populations. 
%Figure~\ref{gcfig}. 
%
%\begin{figure}
%\centering
%\includegraphics[width=0.4\textwidth]{zoo3.eps}
%\includegraphics[width=0.4\textwidth]{GC_SFH.eps}
%\caption{\label{gcfig} 
%Zoom on a giant clumps, showing very dense gas clouds of 1E5-1E6 Msun and 10pc diameter. SFH of a clump. Initial burst during clump collapse, lasting 30-40Myr. Then sustained SFR because of re-accretion of gas, with some peaks (note Figure 7 was smoother over 40Myr bins for readibility)
%}
%\end{figure}

% - - - - - - - - - - - - - - - - - - - - - - - - - - - - - - - - - - -

\section{Summary and Implications}

We have presented simulations of high-redshift clumpy disk galaxies with a new implementation of photoionization and radiation pressure feedback, in addition to supernova feedback, and a physically-motivated estimation of the wind mass loading. Previous models had either considered only weak supernovae feedback models, or used strong feedback recipes generating high-velocity outflows with ad-hoc high mass loading factors. This paper has focussed on the properties and evolution of the giant clumps of $10^{8-9}$\,M$_\odot$ typically found in these galaxies, and the associated birth of outflows. The main findings are:

\begin{itemize}

\item The physically-motivated model for photoionization and radiation pressure feedback from Renaud et al. (2013), used either with a simple thermal model for supernovae feedback or the non-thermal model from Teyssier et al. (2013), produces massive gas outflows from the giant clumps, with outflow rates at the scale of individual clumps of the order of the SFR and up to ten times the SFR over the short periods, where the SFR in a clump is 1--2\,M$_\odot$\,yr$^{-1}$. The outflow properties are consistent with observations. Comparison of various models indicate that the outflow rate is not determined solely by radiation pressure, but is also influenced by the energy injection from supernovae.

\item The giant clumps survive for several $10^8$\,yr, until they migrate through torques and dynamical friction and coalesce centrally after a few hundred Myr. Their mass remains about constant in spite of the gaseous outflows. This is because the clumps are wandering in a gas-rich turbulent disk, from which they constantly accrete gas at a rate of a few M$_\odot$\,yr$^{-1}$, compensating for the mass loss. Feedback-driven outflows ensure regulation of the clump mass after periods of high accretion rates.

\item The continuous accretion of gas sustains long-lasting star formation histories with relatively constant SFR. The average age of stars in a giant clump is significantly younger than the age of the clump itself. In addition, the stars that form in a clump gradually leave the clump through tidal stripping, and up to half of the stars formed in-situ at a given time can have left the clump 200\,Myr later. This preserves young stellar populations ($\leq$\,200Myr) even in older clumps. 

\item Bulge growth through instability-driven inflow, clump migration and coalescence, occurs in reasonable proportions on the present long-lived clump models, especially because the mass of clumps is self-regulated. Clump migration brings stellar mass to the central bulge at an average rate of 1-10 solar masses per year, over a timescale of 500\,Myr for a single generation of clumps or more if cosmological accretion helps to re-form clumps. This typically grows a bulge of $10^{9-10}$\,M$_\odot$ for present-day galaxy masses of $10^{10-11}$\,M$_\odot$, which is a reasonable bulge fraction. Furthermore, given the high fraction of gas in the material conveyed by the clumps, a large fraction of the central mass may evolve into a disky pseudo-bulge. As for more massive clumpy galaxies at $z\, \sim \, 2$, many of them evolve into ellipticals by $z\, = \,0$ and the instability-driven and clump-driven spheroid growth may be an important ingredient in this evolution. 

\end{itemize}

The observed clump properties in $z\,=\,2$ galaxies (outflows and stellar content) are fully consistent with this long-lived clump model. Hence the clumps, even if they represent a limited fraction of the total stellar mass, and more generally the associated violent disk instability, can have a major dynamical influence as studied in previous works: migration to the central bulge, fueling of bright AGN phases up to the Eddington rate (Gabor \& Bournaud 2013, inside-out growth of a thick exponential stellar disk (Bournaud et al. 2007), and perhaps also globular cluster formation (Shapiro et al. 2010) and dark matter cusp erosion \citep{inoue2}. Our feedback model shows that observations of outflowing gas and young stellar populations do absolutely not imply that clumps have to be short-lived transient features. However, since the strength of stellar feedback, and in particular the efficiency of photon trapping in the ISM, remains debated, it remains interesting to consider models with extreme strong feedback (Genel et al. 2012, Hopkins et al. 2012) as a potentially viable alternative\footnote{eventually, unambiguous determination of the effects of photon trapping would require full radiative transfer hydrodynamic simulations with sufficient resolution to resolve the size of young ionized regions in dense gas, i.e. parsec-scale resolution}. These models differ by two aspects. First, they assume a very efficient trapping of the photons emitted by young stars, maximizing the feedback efficiency beyond theoretical expectations (Krumholz \& Thompson 2012, Dekel \& Krumholz 2013). Second, they impost the wind velocity to be equal or close to the local escape velocity, which maximizes the mass loading of the winds. Using a different approach to estimate the mass loading of momentum-driven winds from physical considerations (Renaud et al. 2013) we found here that the wind velocity can be higher, reducing the involved gas mass for a given amount of available momentum. Observations of a radial gradient in the stellar age of clumps \citep{forster,guo} suggests that real clumps do live long enough to migrate radially. Given that the gradient in real clump age should be steeper than in stellar age (Fig.~\ref{ages}), these data do actually suggest lifetimes of a few $10^8$\,yr.

The fact that clumps continuously lose mass and re-accrete gas does not mean that they are permanently disrupted and reformed -- as illustrated in our time series, the clumps are persistent structures that are permanently present with a roughly constant mass. When a clump is analyzed at a given time and 100\,Myr later, it still contains about 80\% of its initial baryons. Over a life-time of 300--500\,Myr, a typical clump will lose about 50--65\% of its initial baryonic mass (through gas outflow and dynamical loss of stars), and will re-accrete a comparable amount of gas. The clumps in unstable disks have a wave-like behavior like spiral arms in lower-redshift galaxies, with the interesting analogy that spiral arms have a very weak contrast in the near-infrared and in stellar mass maps, and mostly contain very young stellar populations and unvirialized molecular clouds, but are density waves that have propagated in the disk for timescales of $10^9$\,yr or more, with important dynamical impact (e.g., Puerari et al. 2000). A noticeable difference is that the timescale for mass exchange is longer for the high-redshift giant clumps, and complete renewal of their mass content does not occur faster than clump migration to the galaxy center.

%%%%%%%%%%%%%%%%%%%%%%%%%%%%%%%%%%%%%%%%%%%%%%%%%%

\acknowledgments
We are grateful to Sarah Newman and Reinhard Genzel for useful comments on the properties of observed outflows. The simulations presented in the work were performed at the {\em Tr\`es Grand Centre de Calcul} of CEA under GENCI allocations 2012-GEN 2192 and 2013-GEN2192, and at the LRZ SuperMUC facility under PRACE allocation number 50816. We acknowledge financial support from the EC through an ERC grant StG-257720 (FB, FR, JMG, KK) and the CosmoComp ITN (JMG, FB). AD was supported by ISF grant 24/12, by GIF grant G-1052-104.7/2009, by a DIP grant, by NSF grant AST-1010033, and by the ICORE Program of the PBC and the ISF grant 1829/12.

\end{document}